\documentstyle[aps,preprint,prb]{revtex}
\tightenlines
\begin{document}
\title{An Integrated Approach to the Assessment of Long Range Correlation
in Time Series Data}
\author{Govindan Rangarajan \thanks{Also associated with the
Jawaharlal Nehru Centre for Advanced Scientific Research,
Bangalore, India; e-mail address: rangaraj@math.iisc.ernet.in}}
\address{Department of Mathematics and Centre for Theoretical Studies\\
Indian Institute of Science \\ Bangalore 560 012, India}
\author{Mingzhou Ding\thanks{E-mail address: ding@walt.ccs.fau.edu}}
\address{Center for Complex Systems and Brain Sciences \\
Florida Atlantic University \\ Boca Raton, FL 33431}
%\date{}
\maketitle

\begin{abstract}
To assess whether a given time series can be modeled by a
stochastic process possessing long range correlation one usually
applies one of two types of analysis methods: the spectral method
and the random walk analysis. The first objective of this work is
to show that each one of these methods used alone can be
susceptible to producing false results. We thus advocate an
integrated approach which requires the use of both methods in a
consistent fashion. We provide the theoretical foundation of this
approach and illustrate the main ideas using examples. The second
objective relates to the observation of long range anticorrelation
(Hurst exponent $H<1/2$) in real world time series data. The very
peculiar nature of such processes is emphasized in light of the
stringent condition under which such processes can occur. Using
examples we discuss the possible factors that could contribute to
the false claim of long range anticorrelations and demonstrate the
particular importance of the integrated approach in this case.

\end{abstract}
\vskip 15pt

\newpage

\section{Introduction}

Random processes with long range power law correlation have been
observed in a variety of fields including economics, geosciences,
physics and biology
\cite{Mandelbrot,Granger,Cassandro,Richards,Hurst,Matheron,Ding,GR,beran}.
There are roughly two types of tools used in assessing the
presence of such correlations in time series data: spectral domain
methods represented by power spectrum analysis
\cite{Mandelbrot,beran} and random walk methods in the time domain
represented by the rescaled range
analysis\cite{Mandelbrot,beran,other}. Often these two types of
tools are applied singly to a given data set. In Section II of
this paper, we point out the pitfalls of this approach through a
series of examples and advocate an integrated approach which
requires the consistent application of both types of methods.

Long range correlations are characterized by a quantity called the
Hurst exponent $H$. When $H>1/2$ the process is said to have
positive long range correlation or persistence while $H<1/2$ means
the process has long range anticorrelation or antipersistence.
When $H=1/2$ we say that the process has short range correlation.
In Section III of this paper we present the condition for $H<1/2$
and discuss how this stringent condition can be corrupted in real
world data. We then proceed to demonstrate the significance of the
integrated approach in this case and analyze the reason underlying
the many reported examples of $H<1/2$.

\section{An Integrated Approach to the Assessment of Long
Range Correlation in Time Series}

In this section, we argue that an integrated approach is required
to assess long range correlations in times series. First, we
derive the relationships between the power law exponents obtained
from analyzing the time series using three different tools --
spectral, autocorrelation and rescaled range analyses. Then we
demonstrate through examples that the use of spectral method or
the rescaled range method alone can produce erroneous results. The
autocorrelation method is not considered since it is often
difficult to use in practice.

\subsection{Theoretical Considerations}

Consider a stationary stochastic process in discrete time,
$\{\xi_k\}$, with $<{\xi}_k>=0$ and $<~{\xi}^2_k>=\sigma^2$. Here
$<>$ denotes ensemble average. If the autocorrelation function
$C(n)=<{\xi}_k{\xi}_{k+n}>$ scales with the lag $n$ as
\begin{equation}
C(n) \sim n^{-\beta} \label{eq:scaling}
\end{equation}
for large $n$, where $0<\beta<1$, then $\{\xi_i\}$ is called a
long range correlated or long memory process \cite{beran}. The
reason for the latter term is that $C(n)$ decays so slowly that
$\sum_{n=1}^{N}C(n)$ diverges as $N \rightarrow \infty$. (The case
of $\beta>1$ will be treated in Section III of this paper.)

The correlation structure of $\{\xi_k\}$ can be conveniently
measured by the power spectrum which is defined as
\cite{chatfield}
\begin{equation}
S(f)=C(0)+2\sum_{n=1}^{\infty}C(n)\cos(2\pi f n). \label{eq:power}
\end{equation}
If $C(n)$ obeys the scaling relation in Eq.~(\ref{eq:scaling}) then
\begin{equation}
S(f) \approx 2\sum_{n=1}^{\infty}n^{-\beta} \cos(2\pi f n). \label{eq:power2}
\end{equation}
In this case, we show below that
\begin{equation}
S(f) \sim f^{-\alpha} \label{eq:sscaling}
\end{equation}
for small $f$ where $\alpha=1-\beta$.

The proof of this relationship between $\alpha$ and $\beta$ draws
on well known results in trigonometric series theory
\cite{zygmund}. Consider the Taylor expansion of the function
$(1-y)^{-\delta-1}$ \begin{equation}
(1-y)^{-\delta-1}=\sum_{n=0}^{\infty}A_n^{\delta}y^n,
\label{eq:expansion}
\end{equation}
where by definition we have $A_0^{\delta}=1$ and for $n \geq 1$
\begin{eqnarray*}
A_n^{\delta}&=&\frac{(\delta+1)(\delta+2)\cdots(\delta+n)}{n!} \\
            &\approx& \frac{n^{\delta}}{\Gamma(\delta+1)}.
\end{eqnarray*}
This means
\begin{equation}
\sum_{n=1}^{\infty}n^{\delta}y^n \approx
\Gamma(\delta+1)[(1-y)^{-\delta-1}-1].
\end{equation}
Replacing $\delta=-\beta$, $y=re^{i2\pi f}$, $0 \leq r<1$, in the
above equation leads to
\begin{equation}
\sum_{n=1}^{\infty}n^{-\beta}r^ne^{i2\pi n f} \approx
\Gamma(1-\beta)[(1-re^{i2\pi f})^{\beta-1}-1].
\end{equation}
Letting $r \rightarrow 1$, $f \rightarrow 0$ and taking the real
part we get
\begin{equation}
\sum_{n=1}^{\infty}n^{-\beta}\cos{2\pi n f} \approx
\Gamma(1-\beta)(2\pi f)^{\beta-1} \cos(\pi(1-\beta)/2).
\end{equation}
Substituting this in Eq. (\ref{eq:power2}) yields
\begin{eqnarray*}
S(f)&\approx&
2\Gamma(1-\beta)(2\pi f)^{\beta-1} \cos(\pi(1-\beta)/2)\\ &\sim&
f^{\beta-1}.
\end{eqnarray*}
Comparing the above with Eq.~(\ref{eq:sscaling}) we obtain
$\alpha=1-\beta$.

Another way to assess the correlation structure of $\{\xi_k\}$ is
to convert the stationary process to a random walk by using
partial sums, $R_1=\xi_1$, $R_2=\xi_1+\xi_2$, $\cdots$,
$R_n=\xi_1+\xi_2+ \cdots+\xi_n$, $\cdots$, where $R_n$ is the
position of the walker at time $n$. The mean range of the random
walk trajectory as a function of time bears specific relations
with the scaling relation Eq.~(\ref{eq:scaling}). For ease of
analytical evaluation we consider the mean square displacement as
a measure of the range of the random walk, which is defined as
\begin{eqnarray}
<R_n^2>&=&\sum_{i=1}^n<\xi_i^2>+2\sum_{s=1}^{n-1}(n-s)C(s)
\nonumber \\
&=&n\sigma^2+2n\sum_{s=1}^{n-1}C(s)-2\sum_{s=1}^{n-1}s C(s).
\label{eq:msd}
\end{eqnarray}
Let $C(s)$ obey the scaling law in Eq.~(\ref{eq:scaling}). The
sums in the above equation are estimated as
\begin{equation}
\sum_{s=1}^{n-1}C(s) \sim \sum_{s=1}^{n} s^{-\beta} \sim \int_1^n
s^{-\beta} \sim n^{1-\beta}
\end{equation}
and
\begin{equation}
\sum_{s=1}^{n-1}s C(s) \sim \sum_{s=1}^{n} s^{1-\beta} \sim
\int_1^n s^{1-\beta} \approx n^{2-\beta}.
\end{equation}
For $0<\beta<1$ this means
\begin{equation}
<R_n^2> \sim n^{2-\beta},
\end{equation}
for large $n$. Conventionally, the mean square displacement is
characterized by the Hurst exponent $H$ as
\begin{equation}
<R_n^2> \sim n^{2H},
\end{equation}
where
\begin{equation}
H=(2-\beta)/2=(1+\alpha)/2. \label{eq:rela}
\end{equation}
Thus we obtain a set of consistent relations between the scaling
exponents $\alpha$, $\beta$, and $H$.

\subsection{Examples}

As shown in the last section the scaling exponents obtained from
spectral analysis and from random walk analysis must be consistent
through Eq.~(\ref{eq:rela}). We first provide three examples of
two simulated and one experimental long range correlated process
demonstrating this consistency. Then we proceed to show that the
spectral method or the random walk method used alone can be
susceptible to artifacts in the data and produce erroneous
results. The combination of the two methods can often detect such
artifacts through inconsistencies with Eq.~(\ref{eq:rela}).

The random walk analysis tool that we will use in this paper is
the rescaled range analysis \cite{Mandelbrot,Hurst}. A brief
description of this method follows. For a given data set $\{ \xi_i
\}$, consider the sum $L(n,s) = \sum_{i=1}^s \xi_{n+i}$, where
$L(n,s)$ can be regarded as the position of a random walk after
$s$ steps. Define the trend-corrected range $R(n,s)$ of the random
walk as
\begin{equation}
R(n,s) = \mbox{max}\{L(n,p)-p L(n,s)/s, \ 1 \leq p \leq s\} -
\mbox{min}\{L(n,p)-p L(n,s)/s, \ 1 \leq p \leq s\}.
\end{equation}
Let $S^2(n,s)$ denote the variance of the data set $\{ \xi_{n+i}
\}_{i=1}^s$. If the data has long range correlation, the average
rescaled statistic $Q(s) = < R(n,s)/S(n,s) >_n$ (where $<>_n$
denotes the average over $n$) scales with $s$ as a power law for
large $s$:
\begin{equation}
Q(s) \sim s^H,
\end{equation}
where $H$ is the Hurst exponent introduced earlier. This power law
manifests itself as a straight line in the log-log plot of $Q(s)$
versus $s$. Spectral analysis was done using Fast Fourier
Transform and the Bartlett window was employed \cite{numerical}.

\subsubsection{Genuine Long Range Correlated Processes} \label{fBm}

To generate a process whose spectral density scales with the
frequency $f$ as a power law, $f^{-\alpha}$, we start with a
realization of a discrete zero mean white Gaussian noise process
$\{ \xi_k \}$, $k=0,1,\dots,N-1$, with variance $\sigma^2$. Using
Fourier Transform we obtain
\begin{equation}
\Gamma_k = \sum_{n=0}^{N-1} \xi_n \exp(-i 2 \pi nk/N), \ \ \
k=0,1,\dots,N-1.
\end{equation}
Next we multiply $\Gamma_k$ by the factor
$f^{-\alpha/2}=(k/N)^{-\alpha/2}$ to obtain the scaled quantity
$\Gamma'_k$. (Both $\Gamma_k$ and $\Gamma_{N-k}$ are multiplied by
the same factor since we want a real-valued time series.) Finally
we perform an inverse Fourier Transform to obtain:
\begin{equation}
x_n = \frac{1}{N} \sum_{k=0}^{N-1} \Gamma'_k \exp(2 \pi  nk/N), \
\ \ n=0,1,\dots,N-1.
\end{equation}
The discrete process ${x_n}$, by construction, has a mean power
spectrum that scales as $f^{-\alpha}$ with the frequency. The
variability around the mean spectrum is provided by the white
noise process.

For our first example we generated a long range correlated process
using the above construction with $\alpha=0.6$ and variance 0.25.
The first data set has 8192 points. When we apply rescaled range
analysis to this data, we obtain a Hurst exponent $H=0.74$ (see
Figure 1(a)). The power spectrum exhibits a power law behaviour
(by construction) with $\alpha=0.58$ (see Figure 1(b)). Note that
these values are consistent with Eq.~(\ref{eq:rela})
\cite{comment1}.

Next, we truncated the above time series data to obtain a short
data set with only 256 points. The results from rescaled range and
power spectrum analyses are shown in Figures 2(a) and 2(b)
respectively. Again these two results are consistent with one
another. This illustrates the fact that when you have a process
with genuine long range correlation, even a short data set is
often sufficient to reveal this property.

As a second example, we consider a different type of long range
correlated process -- a fractional ARIMA(0,d,0) process
\cite{beran} with $0 \le d < 0.5$. It can be shown that the
autocorrelation function $C(n)$ for this process scales with lag
$n$ as $C(n) \sim n^{2d-1}$. Thus from Eq. (\ref{eq:rela}),
$\beta=1-2d$ and $H=d+0.5$. The fractional ARIMA(0,d,0) process
for $d=0.25$ ($H=0.75$) was generated \cite{davies} using its
known autocovariance function \cite{beran}. The results from
rescaled range analysis and spectral analysis of this data, shown
in Figures 3(a) and 3(b) are mutually consistent with one another
[cf. Eq. (\ref{eq:rela})].

The final example analyzes the data from a finger tapping
experiment involving the human sensorimotor coordination
\cite{Ding}. In this experiment, subjects cyclically tapped their
index finger against a computer key in synchrony with a periodic
series of auditory beeps, delivered through a headphone. The data
collected were the synchronization or tapping errors defined as
the time between the computer recorded response time and the
metronome onset (see \cite{Ding} for further details). Here we
analyze the synchronization error time series from this experiment
using rescaled range and spectral analyses. The results, exhibited
in Figures 4(a) and 4(b), are again mutually consistent
demonstrating that the error time series has long range
correlations \cite{Ding}.

\subsubsection{Failures of the Rescaled Range Analysis }

In this section, we consider various situations where rescaled
range analysis used alone can give wrong results.

As our first example, we consider the superposition of an
exponential trend over a white noise process. (See [11] for more
examples in this area.) Specifically, we generated the following
discrete process:
\begin{equation}
x_k = \exp(-0.01k)+\xi_k,\ \ \ k=1,2,\dots, \label{eq:exp}
\end{equation}
where $\{\xi_k\}$ is a white noise process with zero mean and
variance 0.16. A total of 8192 points were generated. This example
can be realized in situations where the process under
investigation has an exponentially decaying transient and one does
not discard the initial portion of the data (containing this
transient) while recording it.

When the above process is subject to rescaled range analysis, we
obtain a Hurst exponent equal to 0.75 (see Figure 5(a)) indicating
long range correlation where there is none. On the other hand, the
power spectrum is flat (see Figure 5(b)) and does not show any
power law behaviour. This inconsistency between the spectral
method and rescaled range method serves as a warning sign pointing
to the need for further more careful examination of the data.

As our second example, we consider the following autoregreesive
process of order one [AR(1)] \cite{chatfield}:
\begin{equation}
x_k = \lambda x_{k-1} + \xi_k,\ \ \ k=1,2,\dots,
\label{eq:ar1}\end{equation} where $\{\xi_k\}$ is a zero mean
white noise process with variance 0.25 and the coefficient
$\lambda$ is close to 1 (0.9 in our case). The autocorrelation
function of the $x$ process decays exponentially $C(k)\sim
{\lambda}^k$. This means there is no long range correlation in the
$x$ process. However, as shown in Figure 6(a), the rescaled range
analysis of the above process (with 1024 points) indicates the
presence of long range correlation by producing $H=0.76$. The
power spectrum (Figure 6(b)), on the other hand, exhibits a
flattening trend at low frequencies and contradicts the result
from the rescaled range analysis. Even if one misses this
flattening trend and fits a straight line to the remaining portion
of the spectrum on a log-log scale, we get a value for $\alpha$
equal to -1. Here the consistency relation $H \approx
(1+\alpha)/2$ is not satisfied, thus indicating the absence of
long range correlations.

As our third example, we again consider an AR(1) process [cf. Eq.
(\ref{eq:ar1})] but this time with the coefficient $\lambda$ close
to -1 (-0.9 in our case). In this example, the application of the
rescaled range analysis gives a Hurst exponent $H=0.33$ (Figure
7(a)). Naively, this Hurst value can be interpreted as an
indicator of long range ``anti-persistence'' \cite{Mandelbrot}. As
before, the power spectrum contradicts this result (see Figure 7(b))
and shows a flattening trend in the low frequencies.
This observation is further strengthened by analysing a long
data set (100,000 points) using rescaled range analysis. The
results (see Figure 7(c)) show that $H$ approaches a value
of 0.5 as the data set gets longer. In Section
III of this paper we will discuss in more detail processes with
Hurst exponent $H<1/2$.

We would like to make one point regarding the application of the
surrogate data analysis \cite{theiler} which is often used in
combination with many analysis methods to strengthen their results
by demonstrating that a completely random process could not have
exhibited the observed results. We show below that this is not
fool proof when used in conjunction with rescaled range analysis.
We have already seen in the second example above that rescaled
range analysis indicates the presence of long range correlation in
AR(1) process (with $\lambda=0.9$) where there is none. We now
apply surrogate data analysis by shuffling the data randomly and
reapplying rescaled range analysis to the shuffled data sets.
Figure 8 shows the comparison between the Hurst exponent obtained
from the unshuffled original data with the average value of Hurst
exponents obtained from five realizations of randomly shuffled
data. We see that the shuffled data gives an average value of $H$
around 0.5 as compared to 0.76 for the original data. The two
results are well separated. Therefore, the application of
surrogate data analysis would indicate that the result obtained by
rescaled range analysis of the original data is correct,
indicating the presence of long range correlation. This is
obviously a false conclusion. This example demonstrates that
surrogate data analysis can not be used indiscriminately for this
type of problems.

\subsubsection{Failure of the Power Spectrum Analysis}

Here we give an example where the use of power spectrum analysis
with inappropriate parameters can lead to wrong results. If we
investigate the genuine long range process introduced above using
the power spectrum analysis with Parzen window and $M=20$ (that
is, with a lot of smoothing)\cite{chatfield}, then we obtain the
spectrum given in Figure 9. This power spectrum shows a flat
portion at low frequencies indicating wrongly the absence of long
range correlations. This is not a problem with power spectrum
analysis per se, but is an example of using it with inappropriate
parameters. We do not run into such problems with rescaled range
analysis since it does not have such free parameters that can be
wrongly ``tuned''. The above example is not as artificial as it
seems since canned power spectrum analysis routines are often used
in data analysis without proper thought going into the choice of
input parameters. In this case the inconsistencies between the two
analysis methods will prompt more careful examinations of the
methods employed.

\subsubsection{Failures of the Combined Use of Rescaled Range and
Power Spectrum Analysis}

All the above examples illustrate the fact that one should not
rely on a single tool to analyze time series data. An integrated
approach requiring the consistent use of several available tools
is more desirable. But even an integrated approach is not
foolproof as we show below.

Consider a process that is the superposition of AR(1) processes.
In particular, we consider a variable that is the sum of following
five independent processes:
\begin{equation}\label{5ar1}
x_k = \lambda x_{k-1} + \xi_k,\ \ \ k=1,2,\dots,
\end{equation}
where the coefficient $\lambda$ for the individual processes are
given by 0.99, 0.9, 0.4, 0.2, 0.1 and the variances are given by
0.05, 0.1, 0.3, 0.4, and 0.5 respectively. In this case,
both the rescaled range and power spectrum analyses indicate the
presence of long range correlation in the data (see Figures 10(a) and
10(b) respectively) by yielding $H=0.96$ and $\alpha=0.82$. Clearly,
the results are consistent with $H \approx (1+\alpha)/2$.

This example shows that even the superposition of a few AR(1)
processes can mimic a long range correlated process for short data
sets. Theoretically it is known that the superposition of an
infinite number of AR(1) processes can, in some cases, give rise
to a long range correlated process\cite{Granger2}. Even though an
integrated approach using both rescaled range and power spectrum
analyses can give spurious results, consistent positive results
from both these analyses at least indicates the presence of
multiple time scales in the data set.

\section{The Condition for $H<1/2$ and Its Implications for Real
World Data Analysis}

As mentioned earlier, rescaled range analysis
\cite{Hurst,Mandelbrot} is often used in determining the presence
of long range correlation in data sets. Results of rescaled range
analyses are typically quantified using the Hurst exponent $H$ ($
0 < H < 1$). In principle, analysis of a data set can lead to any
value of $H$ between 0 and 1. In this section, we will argue that
processes with $H<1/2$ are rather special in that they must
satisfy the condition that the sum of the autocorrelation function
be zero. Many physical processes are known to meet this condition
\cite{Richards}. However, this condition can be easily corrupted
in real world data where noise unrelated to the physical process
enters the measurement. We demonstrate the importance of the
integrated approach in the proper diagnosis of processes with
$H<1/2$. We also discuss a common way in which a misjudgment of
long range anticorrelation can occur. In this regard we identify
the three contributing factors:  (1) the variable being recorded
is not fundamental (see below), (2) the data set is short and (3)
only random walk type of analysis is employed.

\subsection{The Condition for $H < 1/2$} \label{theory}

Refer to Eq.~(\ref{eq:msd}). Suppose that $C(s) \sim s^{-\beta}$.
When $0<\beta<1$, we showed that both the second and the third
term in the above equation diverge and scale with $n$ as $\sim
n^{2-\beta}$. Therefore $<R_n^2>$ scales with $n$ as $\sim
n^{2-\beta}$ for large $n$ and this leads to $H>1/2$ [cf. Eq.
(\ref{eq:rela})]. For $\beta>2$ (that is, for any $C(s)$ that
decays faster than $C(s) \sim s^{-2}$), both sums in the above
equation converge and we generally obtain $H=1/2$.

Thus the only remaining range of $\beta$ is $1<\beta<2$. For such
$\beta$ the sum $\sum_{s=1}^{\infty}C(s)$ is finite. (Note that,
strictly speaking, this type of process can no longer be termed a
long memory process based on the definition in Section II A. But,
since it has the potential of giving $H<1/2$ we will still use the
term long range correlation.) Therefore the first two terms scale
as $\sim n^1$. The sum $\sum_{s=1}^{n-1}s C(s)$ in the third term
is evaluated to be
\begin{equation}
\sum_{s=1}^{n-1}s C(s)\sim n^{2-\beta} \label{eq:div}
\end{equation}
where $(2-\beta)<1$. This means that the rate of divergence of the
third term is slower than the first two terms. Therefore, in the
large $n$ limit, $<R_n^2> \sim n$ we would still observe $H=1/2$.
The only situation when this will not happen occurs when the
following equation is precisely satisfied
\begin{equation}
\sigma^2+2\sum_{s=1}^{\infty}C(s)=\sum_{s=-\infty}^{s=\infty} C(s)=0
\label{eq:condition}
\end{equation}
In this case the first two terms in $<R_n^2$ drop out giving
$<R_n^2 \sim n^{2-\beta}$. Therefore
we obtain $H=(2-\beta)/2$ which is smaller than 1/2.

It is clear that for $H<1/2$ to occur the process must meet a very
stringent condition, Eq.~(\ref{eq:condition}). It has been shown
that many physical systems satisfy this condition \cite{Richards}.
But, when such a physical system is subject to measurement, noise
is an inevitable factor. For the noisy measurement it is likely
that the equality in Eq.~(\ref{eq:condition}) no longer holds. The
implication is that in the long run one may observe $H=1/2$ and
therefore not be able to correctly identify the underlying
process. In what follows we show that the integrated approach
advocated in the previous section is again an essential tool in
revealing strong clues as to the true nature of the physical
process. Moreover, we will show that the integrated approach is
also indispensable in guarding against misjudgment of $H<1/2$ for
systems where this is not true.

\subsection{Integrated Approach to Noisy $H<1/2$ Data}

For a genuine $H < 1/2$ process, Eq.~(\ref{eq:rela}) still holds,
albeit $\alpha$ will be a negative number. We generated a
$f^{-\alpha}$ process artificially using the procedure in
Section~\ref{fBm}. A value of $\alpha =-0.5$ was used. When the
data is subject to rescaled range analysis, we obtain a value of
$H=0.28$ (see Figure 11a). The spectral analysis gives a power law
curve with $\alpha = -0.5$ as expected (Figure 11b). We note the
results of the rescaled range and spectral analyses are mutually
consistent in this case since the application of Eq.
(\ref{eq:rela}) gives a $H$ value of 0.25.

Now we consider the effect of additive noise on the same data set.
When noise is added, Eq. (\ref{eq:condition}) is no longer strictly
satisfied. Hence we would expect the Hurst exponent of the process to
asymptotically approach $H=0.5$ for long data sets. This is borne
out by our numerical simulations. We start with the true long range
correlated process described above ($H=0.25$) and add Gaussian white
noise to the data with variance 0.01. This simulates the effect
of noisy measurement in experiments. Figure 12(a) shows the results
of rescaled range analysis applied to the above process with
262,144 points. We see that $H$ asymptotically tends to 0.5.
As the variance of the added noise is increased, the
$H=0.5$ value is reached even faster. However,
all is not lost. The integrated approach allows us to approximately
recover the true process hidden by the noise. We start by truncating
the data set which prevents the asymptotic limit for $H$ being
reached. Figures 12(b) and 12(c) display the results of applying
the integrated approach to the truncated data set with 8192 points
(all other parameter values remain the same as in Figure 12(a)). We
obtain a $H$ value of 0.34 from rescaled range analysis and the
spectral analysis is consistent with this value. This consistency
tells us that the data represents a true long range correlated
process (the $H$ value obtained is higher than that for the true
process because of the added noise.) We comment that this consistency is in
marked contrast to what was observed for the AR(1) process with
$\lambda=-0.9$ (see Section IIB). There, even though the $H$ value
was 0.33 for small data sets and rose to $H=0.5$ for large data
sets, the results of spectral analysis were completely inconsistent
with this. This lead us to conclude that there was no true long
range correlated process in that example. The above examples
again demonstrate that the integrated approach is very useful
in revealing the true nature of a process represented by time
series data.

\subsection{Possible Causes for False Identification of $H < 1/2$}

In the literature one often comes across reports where analysis of
real world data, using the method of random walk alone, yield
$H<1/2$. The examples in Section II B show the shortcoming of
using just one type of analysis method. Upon further examination
we realize that there is a common thread in these reports that has
to do with the fact that the data analyzed does not come from a
fundamental process which we discuss below.

\subsubsection{The Notion of a Fundamental Process and Data
Differencing}

Consider a stationary process $\{\xi_k\}$. By definition a
stationary process is not diffusive. In other words, $<\xi_k^2>$
is a constant. Consider the partial sum $R_n=\sum_{k=1}^n \xi_k$.
If $<R_n^2>$ increases with $n$, namely, $R_n$ is a diffusive
process we say that $\{\xi_k\}$ is a fundamental process. If the
time series data coming from a fundamental process is subject to
random walk type of analysis such as the rescaled range analysis,
it can be trusted to correctly assess the corresponding Hurst exponent.

A differenced process refers to a process $\{\eta_k\}$ which is
obtained by $\eta_k=\xi_{k+1}-\xi_k$. Clearly, $\{\eta_k\}$ is not
a fundamental process since its partial sums give $\{\xi_k\}$
which is not diffusive. This means if we input data from a
differenced process into the rescaled range type of random walk
analysis we will not be able to assess the correlation properties
of the original process, and possibly even be fooled by the
appearance of the rescaled range plot (see below).

Differenced data can arise in practice in a number of ways. First,
differencing is a commonly applied technique for trend removal
\cite{chatfield}. Second, the measured physical variable is a
derivative of another fundamental variable. We believe that the
use of differenced data, in combination with rescaled range type
of analysis, underlies some of the reported cases of $H<1/2$. Below we
demonstrate this point by examples.

\subsubsection{Gaussian White Noise}

Consider a Gaussian white noise process \{$\xi_i$\}. The partial
sums of this process yield a diffusive Brownian motion with a
Hurst exponent equal to 1/2. Suppose what is being measured is not
$\xi_i$ but the differenced variable $y_i = \xi_i - \xi_{i-1}$. To
see the effect of this, a Gaussian white noise process \{$\xi_i$\}
with zero mean and variance 0.25 was first generated. If the
rescaled range analysis is performed on $\{ y_i \}$, the result is
shown in Figure 13(a). The data length is 1024 points. If we force a
linear fit to the end part of the log-log plot we observe a Hurst
exponent equal to about 0.12. But this exponent is not a
reflection of the process but caused by the finite size of the
data set. If we analyze a much longer data set (40000 points), we
observe that the slope of 0.12 that we had seen earlier is only a
transient effect (see Figure 13(b)). It can be shown that the true
slope goes to zero as the time lag $s$ increases.

The $H$ value of 0.12 obtained for the differenced data set of
1024 points can also be easily rejected by the integrated
approach. Subjecting the same data set to the spectral analysis
yields $\alpha = -1.95$ (Figure 13(c)). This value is totally
inconsistent with the $\alpha=-0.76$ predicted by Eq.
(\ref{eq:rela}) based on $H=0.12$. This inconsistency should be
used as a clue to further examine the nature of the data set.

\subsubsection{Langevin Equation}

In this section, we consider a more physical example -- the
Langevin equation. The Langevin process that we studied is given
below
\begin{equation}\label{eq:langevin}
\dot{x} = - \lambda x + \xi(t),
\end{equation}
where $\xi (t)$ is a white noise process with zero mean and
variance 0.25 and $\lambda = 5$. The above stochastic
differential equation was integrated using an efficient
algorithm \cite{rao}.

Suppose that the variable being measured is the velocity $\dot{x}$
and successive values of $\dot{x}$ by the numerical integration
scheme constitutes our data set. The rescaled range analysis
applied to a short data set of 250 points yields a Hurst exponent
equal to 0.18 (see Figure 14(a)). On the other hand, if the size of
the data set is increased to 60,000 points, the value of $H$
becomes nearly zero for large $s$ values (Figure 14(b)).

Theoretically, it can  be shown that $x$, after an initial
transient, is a stationary fundamental process. To demonstrate
this we apply rescaled range analysis to the values of $x$. In
this case, even for short data sets, we get a value of $H$ close
to 0.5 (see Figure 14(c)), as predicted by theory. The result seen
in Figure 14(a) is therefore the consequence of $\dot{x}$ being an
over differenced variable.

\subsubsection{The Finger Tapping Data}

Generally, we suggest that when a $H<1/2$ is obtained from a
random walk type of analysis, one should also perform spectral
analysis on the same data set. If the result is inconsistent with
Eq.~(\ref{eq:rela}) then one should conclude that this data set is
not from a fundamental process and partial sums of this data set
should be considered instead for analysis of the correlation
properties.

As an example, we again consider the finger tapping experiment
\cite{Ding} described in Section IIB. But now we analyse the
interresponse intervals (IRIs) instead of the synchronization
errors. The IRI can be obtained from the synchronization error
data by differencing it \cite{Ding} and is itself an important
physiological variable. Rescaled range analysis of this IRI time
series data appears to give a value of $H=0.25$ (see Figure
15(a)). But this is an artifact of the finite data size. This can
be seen by performing a spectral analysis on the same IRI data
(see Figure 15 (b)). We see that this gives results inconsistent
with Eq.~(\ref{eq:rela}). Thus, the $H$ value obtained in Figure
15(a) is a consequence of the IRI being an over differenced
variable, combined with finite data size and only one type of
method.

\section{Summary}

Suppose that the autocorrelation function $C(s)$ for a stationary
process scales with $s$ as $C(s) \sim s^{-\beta}$. Depending on
the values of $\beta$ and the behavior of
$\sum_{s=-\infty}^{s=\infty}C(s)$ we have the following
classification for the process: (1) if $0<\beta<1$ we have
$1/2<H<1$ and the process is said to have long range persistent
correlation or long memory \cite{beran}, (2) if $1<\beta<2$ and
$\sum_{s=-\infty}^{s=\infty}C(s)=0$ we have $0<H<1/2$ and the
process is said to have long range antipersistent correlation or
anticorrelation, and (3) if $\beta>1$ and
$\sum_{s=-\infty}^{s=\infty}C(s) \neq 0$ we have $H=1/2$ and the
process is said to have short range correlation. It is worth
noting in this classification processes with $1<\beta<2$ can be
classified as either having long range anticorrelation or short
range correlation depending whether
$\sum_{s=-\infty}^{s=\infty}C(s)$ is zero. We preserve the long
range anticorrelation terminology in keeping with the traditional
naming of such processes. The main goal of this work has been to
demonstrate the importance of the integrated approach, combining
both spectral and random walk analysis, to the assessment of
correlation behavior in time series data. We showed that the
consistent use of both spectral and random walk analysis is not
only essential in revealing the true nature of a given process it
can also prevent the false conclusion of long range correlation
resulting from artifacts or wrong measurement variables combined
with just one type of analysis methods.

\section*{Acknowledgements}

This work was supported by US ONR Grant N00014-99-1-0062. GR
thanks Center for Complex Systems and Brain Sciences, Florida
Atlantic University, where this work was performed, for
hospitality. We thank the referee for useful suggestions.

\newpage

\section*{Figure Legends}

\begin{description}

\item{\bf Figure 1:} (a) Log-log plot of the rescaled range statistic $Q(s)$
against window size $s$ for a true long range correlated process with
$\alpha=0.6$,
variance 0.25 and a long data set (8192 points). (b) Spectral density of
the same process.

\item{\bf Figure 2:} (a) Log-log plot of the rescaled range statistic $Q(s)$
against window size $s$ for a true long range correlated process with
$\alpha=0.6$,
variance 0.25 and a short data set (256 points). (b) Spectral density of
the same process.

\item{\bf Figure 3:} (a) Log-log plot of the rescaled range statistic $Q(s)$
against window size $s$ for a fractional ARIMA(0,d,0) process with
$d=0.25$. (b) Spectral density of
the same process.

\item{\bf Figure 4:} (a) Log-log plot of the rescaled range statistic $Q(s)$
against window size $s$ for the synchronization error time series data
from the finger tapping experiment. (b) Spectral density of
the same data.

\item{\bf Figure 5:} (a) Log-log plot of the rescaled range statistic $Q(s)$
against window size $s$ for the superposition of an exponential trend over a
white noise process. (b) Spectral density of the same process.

\item{\bf Figure 6:} (a) Log-log plot of the rescaled range statistic $Q(s)$
against window size $s$ for an AR(1) process with $\lambda=0.9$.
(b) Spectral density of the same process.

\item{\bf Figure 7:} (a) Log-log plot of the rescaled range statistic $Q(s)$
against window size $s$ for an AR(1) process with $\lambda=-0.9$,
variance 0.25 and 1024 data points.
(b) Spectral density of the same process.
(c) Log-log plot of the rescaled range statistic $Q(s)$
against window size $s$ for the same process but with 100,000 points.

\item{\bf Figure 8:} Comparison between the Hurst exponent obtained from
the unshuffled original data of the AR(1) process (with $\lambda=0.9$) and
the average value of Hurst coefficients obtained from five realizations of
the above data randomly shuffled.

\item{\bf Figure 9:} Spectral density of the genuine long range correlated
process considered in Figure 1 using a Parzen window with $M=20$.

\item{\bf Figure 10:} (a) Log-log plot of the rescaled range statistic $Q(s)$
against window size $s$ for a superposition of five AR(1) process.
(b) Spectral density of the same process.

\item{\bf Figure 11:} (a) Log-log plot of the rescaled range statistic
$Q(s)$ against window size $s$ for a true long range correlated
process with $\alpha=-0.5$, variance 0.25 and a long data set
(8192 points). (b) Spectral density of the same process.

\item{\bf Figure 12:} (a) Log-log plot of the rescaled range statistic
$Q(s)$ against window size $s$ for a true long range correlated
process corrupted by added white noise with $\alpha=-0.5$, variance 0.25
and 262,144 points. The  variance of added white noise is 0.01.
(b) Same as above but with 8192 points.
(c) Spectral density of the same process with 8192 points.

\item{\bf Figure 13:} (a) Log-log plot of the rescaled range statistic
$Q(s)$ against window size $s$ for a Gaussian white noise process
with 1024 data points. (b) Same as above but with 40,000 data
points. (c) Spectral density of the process in (a).

\item{\bf Figure 14:} (a) Log-log plot of the rescaled range statistic
$Q(s)$ against window size $s$ for the velocity variable $\dot{x}$
of a Langevin process with $\lambda=5$ and 250 data points. (b)
Same as above but with 60,000 data points. (c) Log-log plot of the
rescaled range statistic $Q(s)$ against window size $s$ for the
position variable $x$ of a Langevin process with $\lambda=5$ and
250 data points.

\item{\bf Figure 15:} (a) Log-log plot of the rescaled range statistic $Q(s)$
against window size $s$ for the interresponse interval time series data
from the finger tapping experiment. (b) Spectral density of
the same data.

\end{description}


\begin{references}
\bibitem{comment1} We comment that consistency here is defined in a rather loose
fashion. In the future we will use more rigorous statistical
criteria to provide the basis for evaluating whether the estimates
from the two methods are consistent with Eq.~(\ref{eq:rela}).
\end{references}
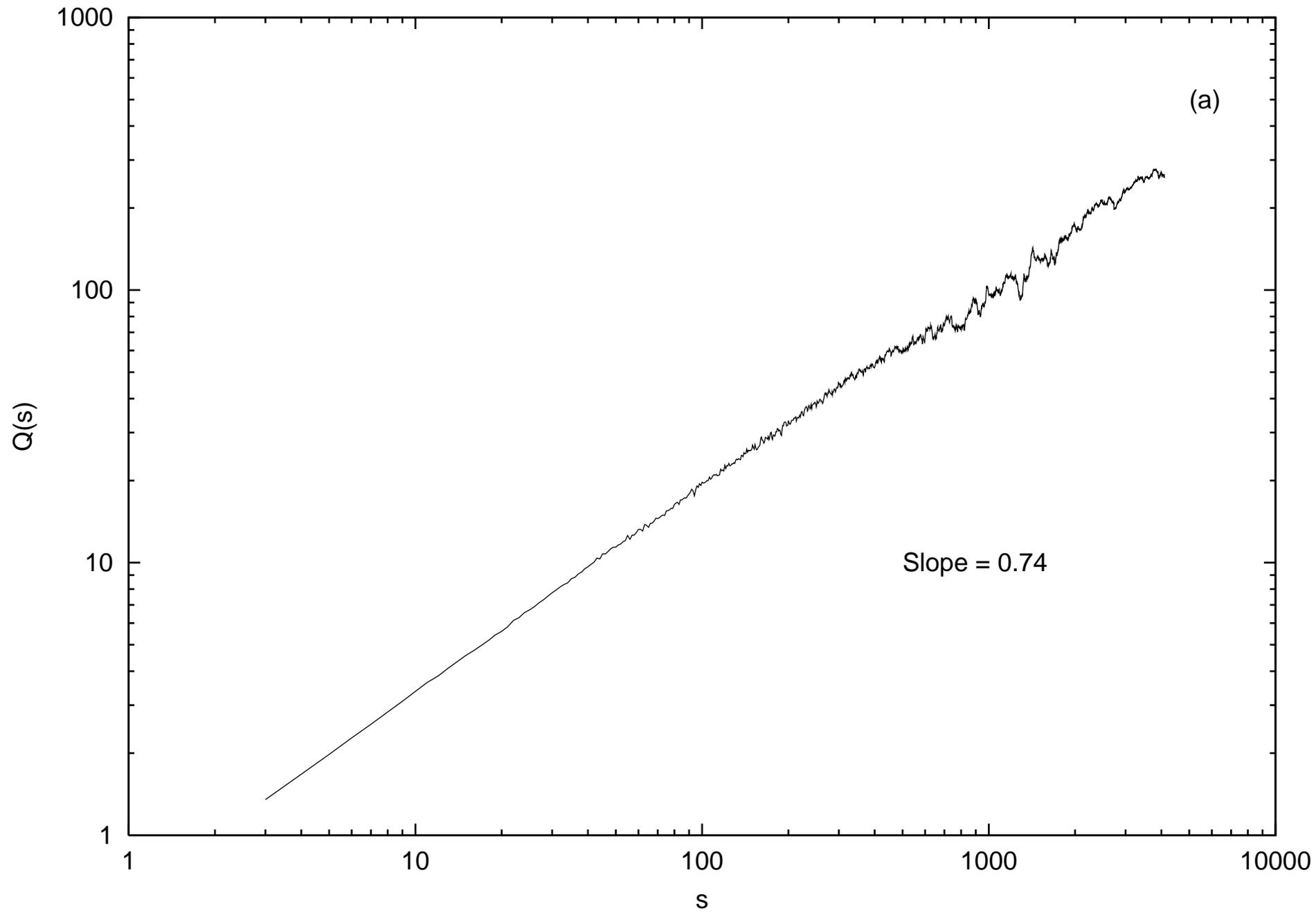
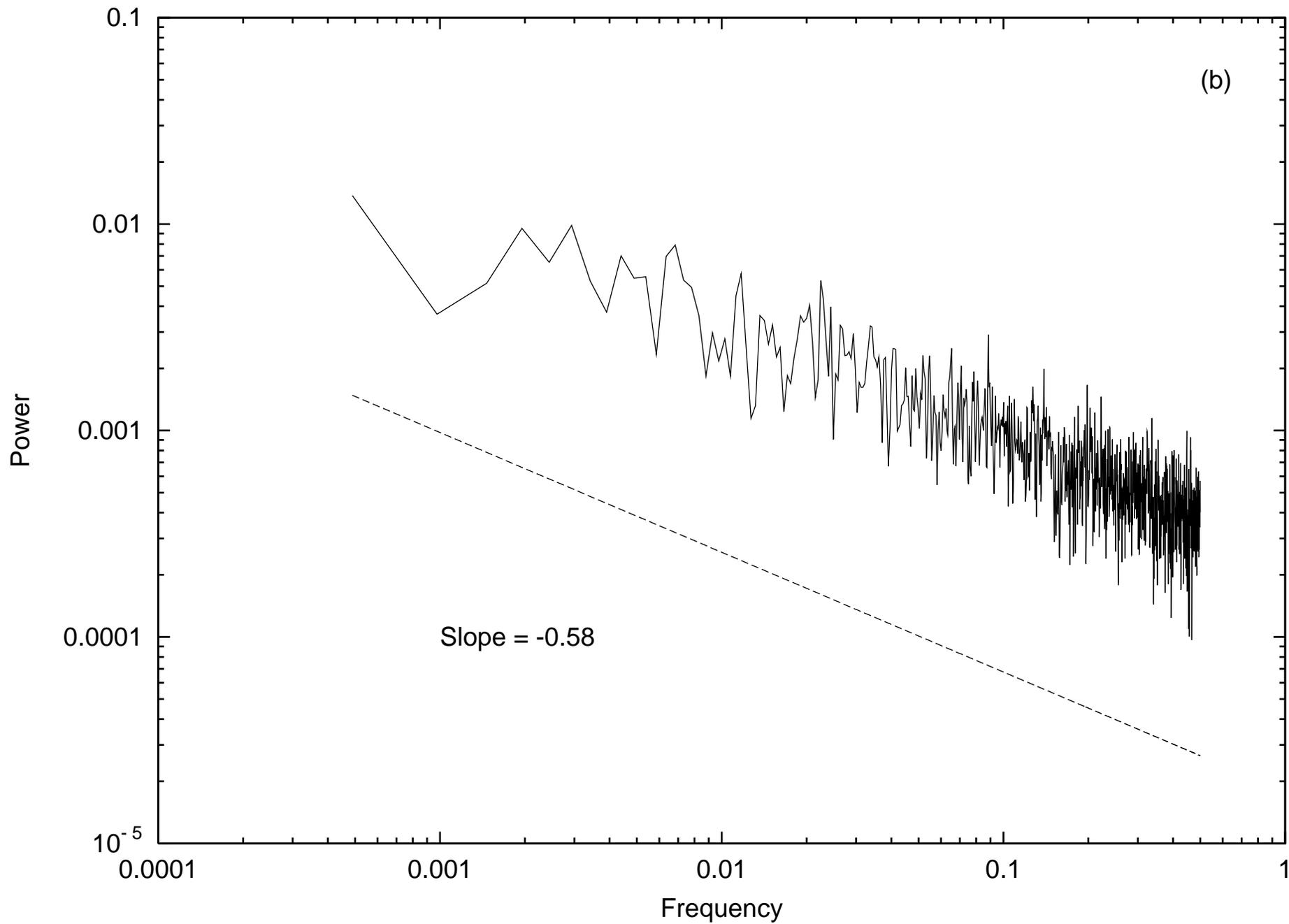
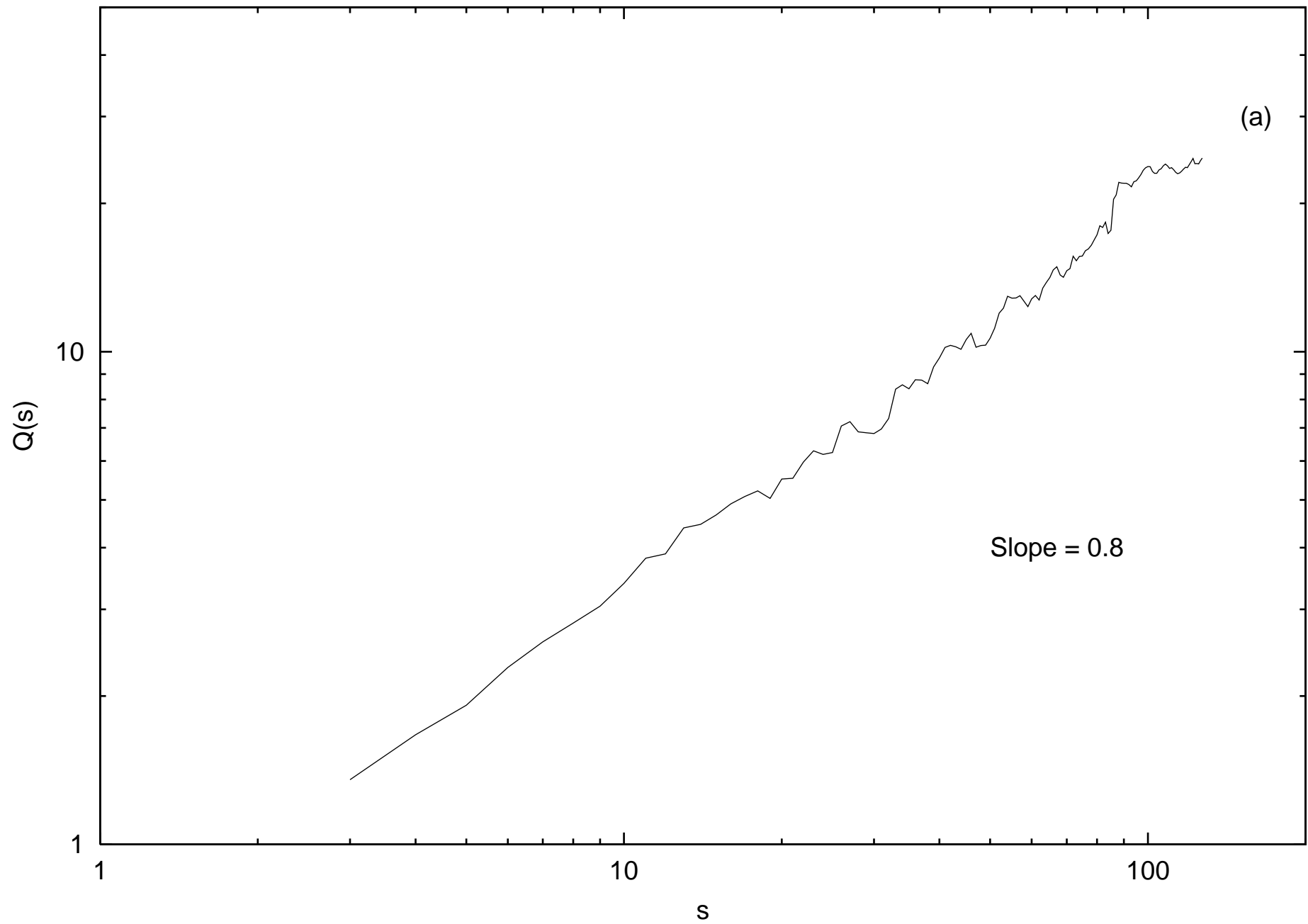
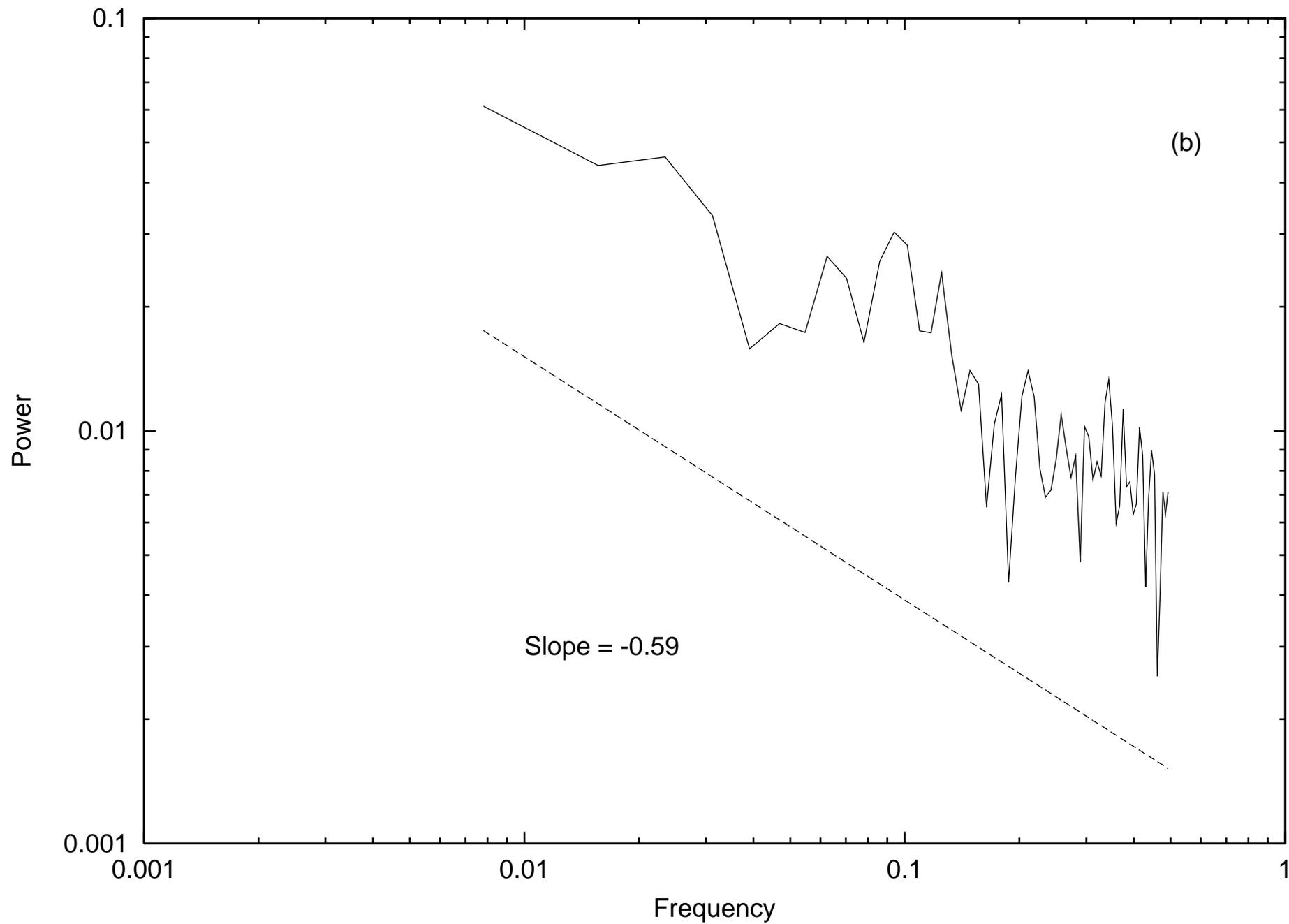
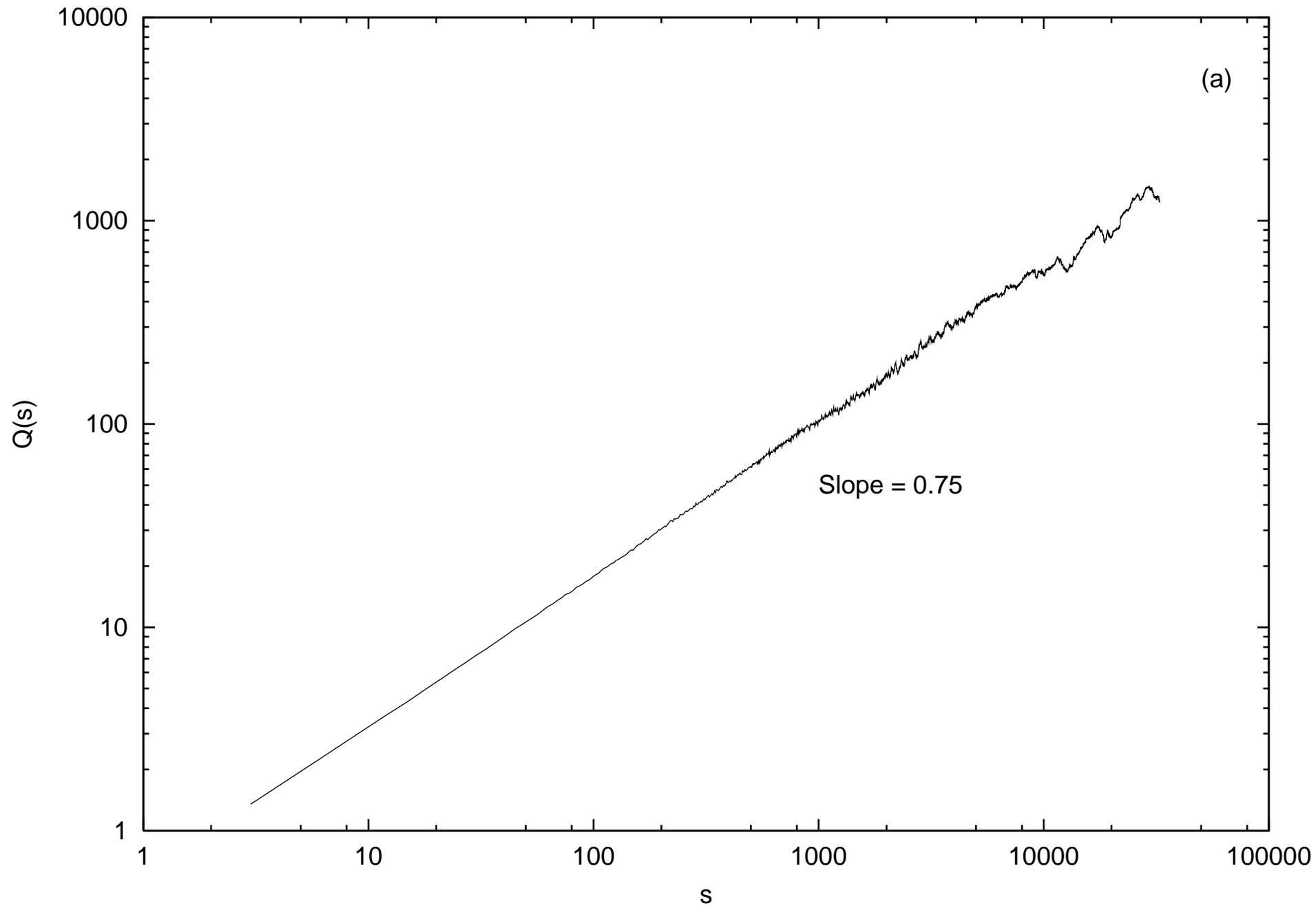
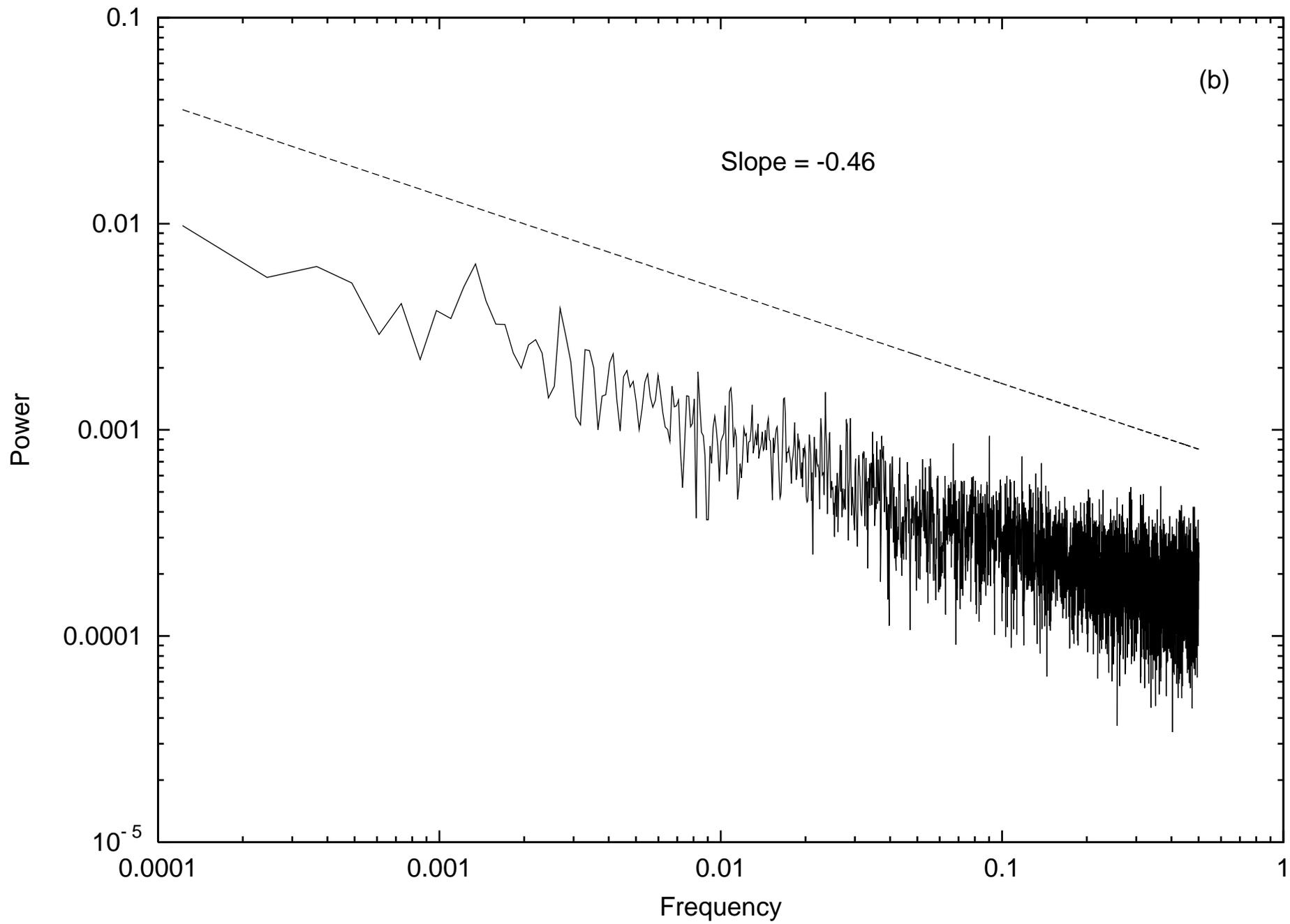
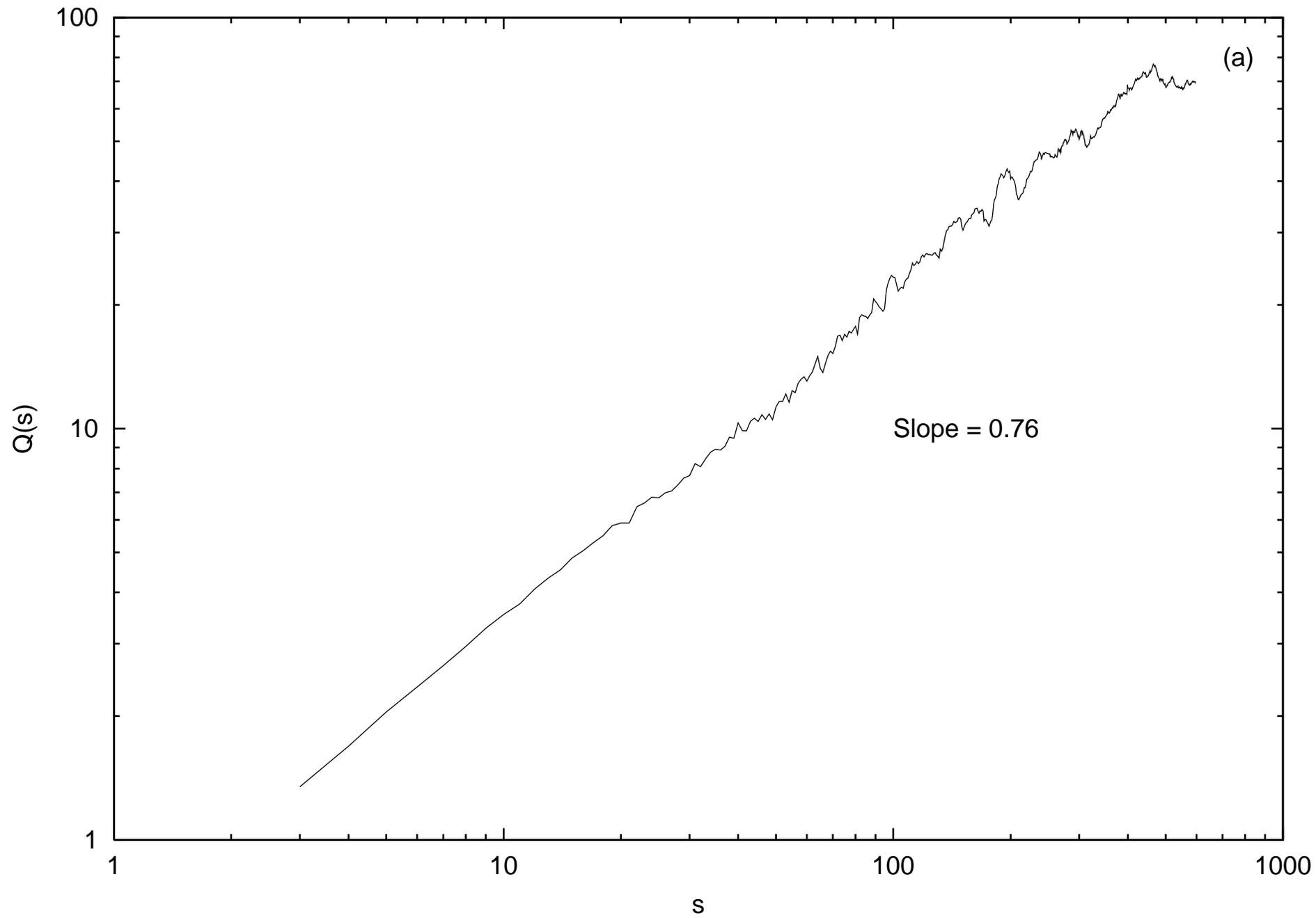
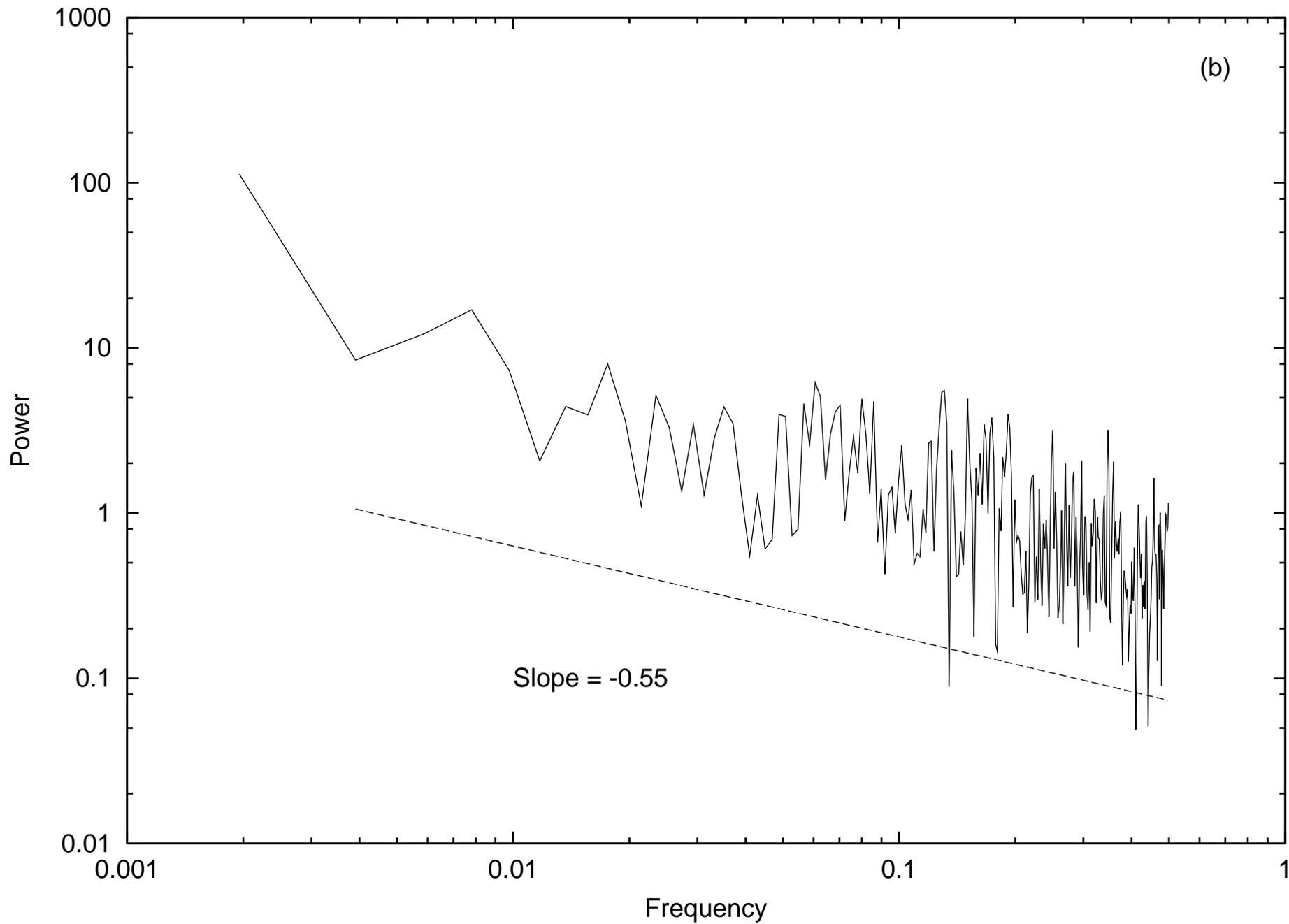
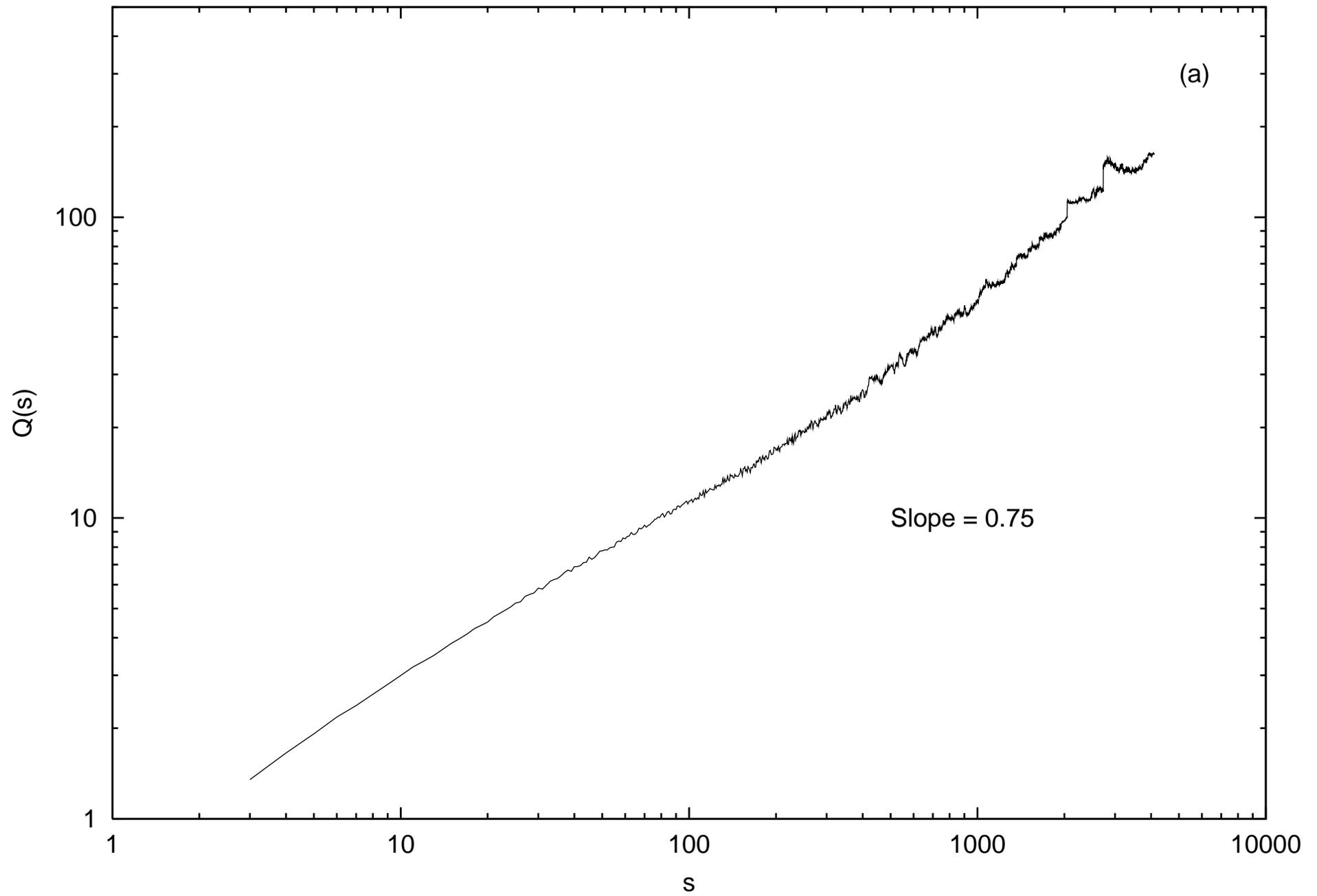
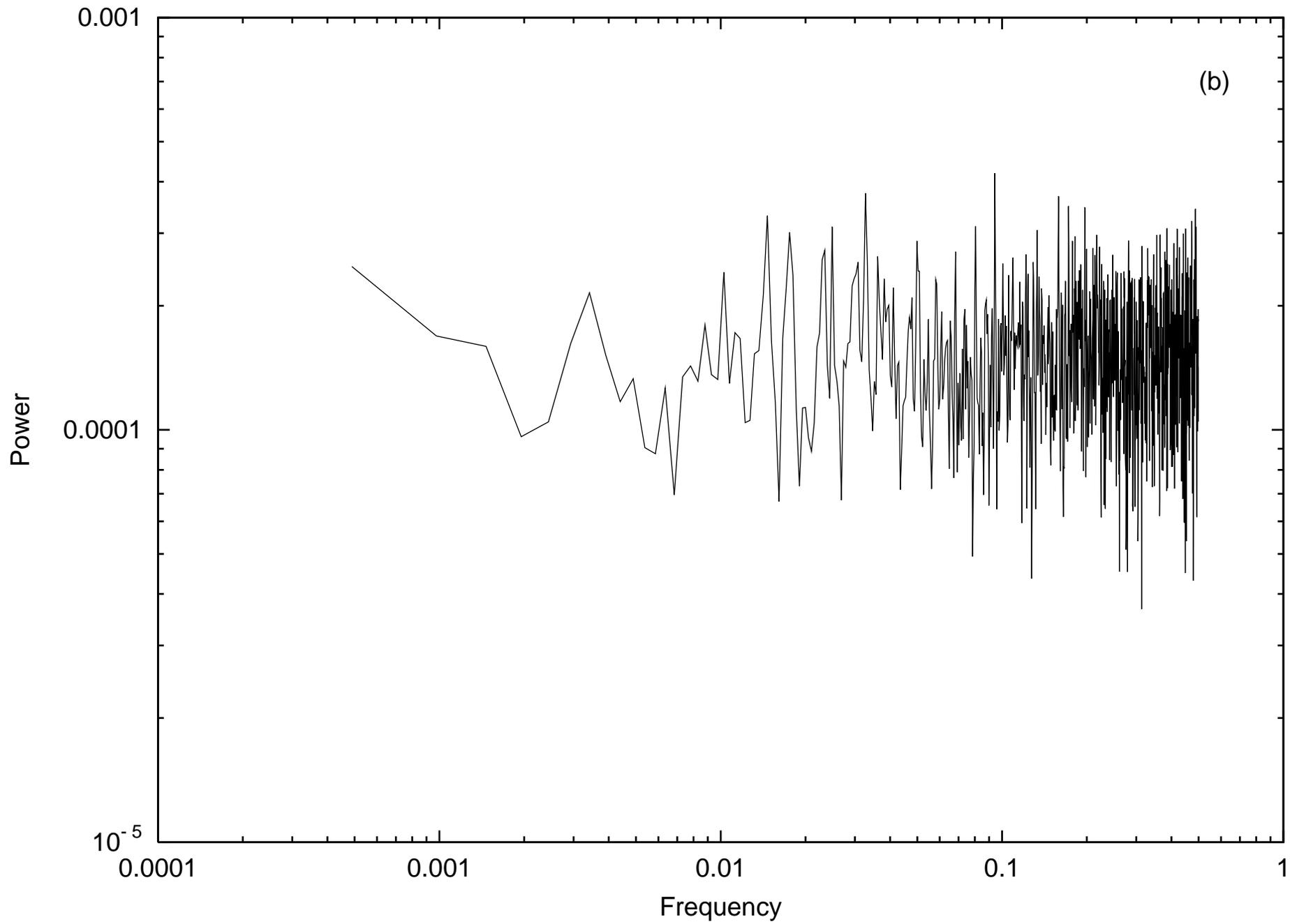
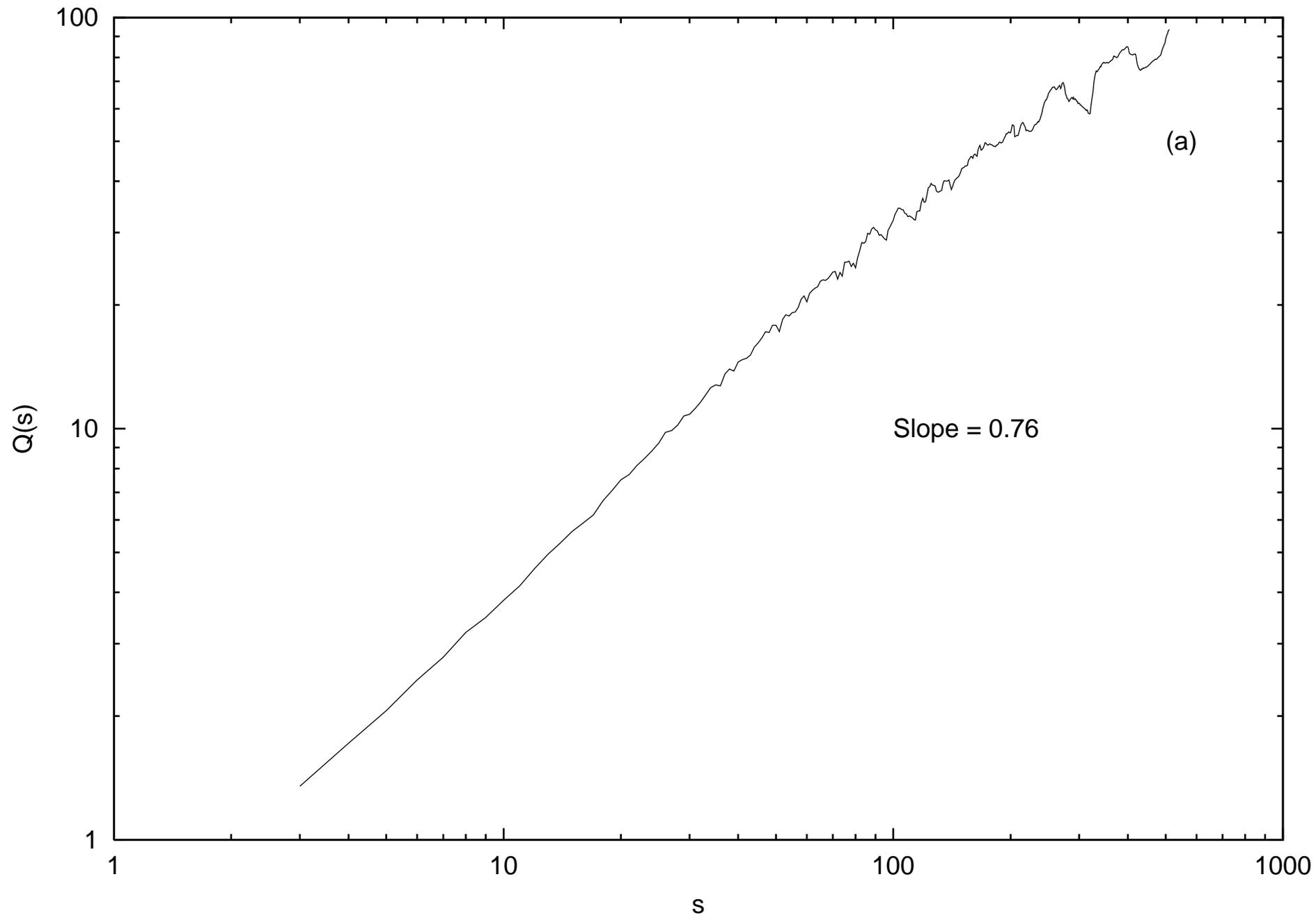
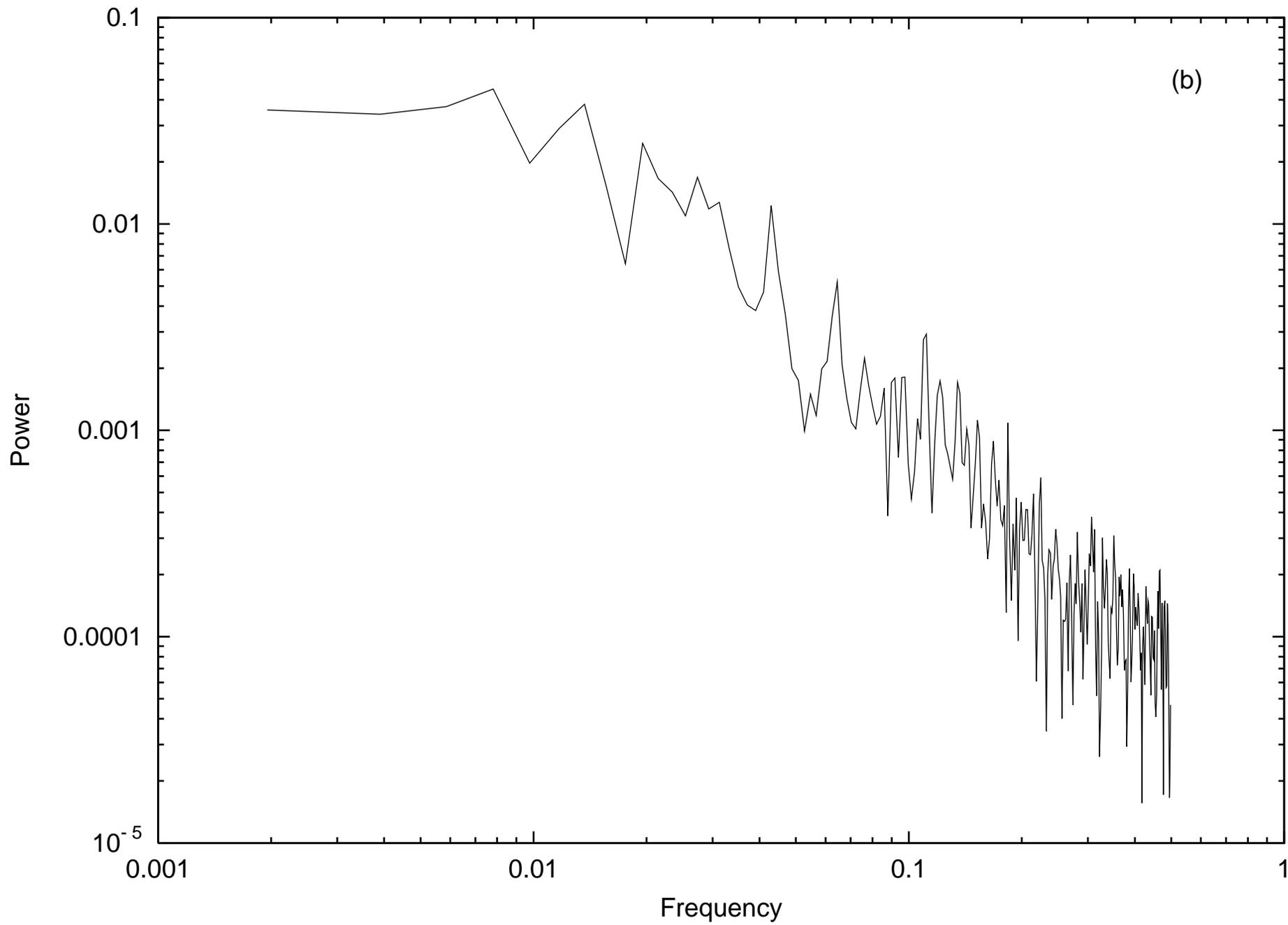
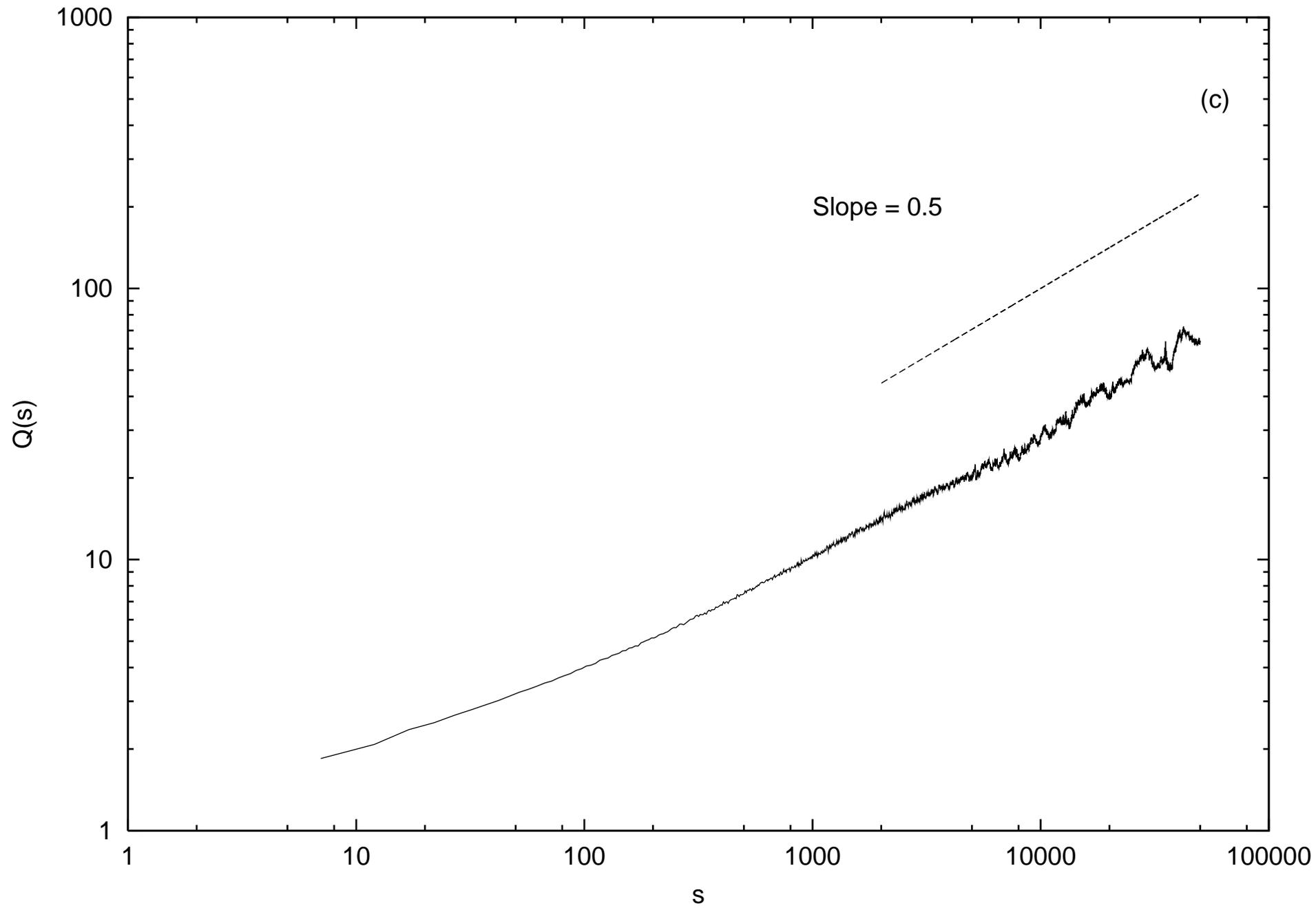
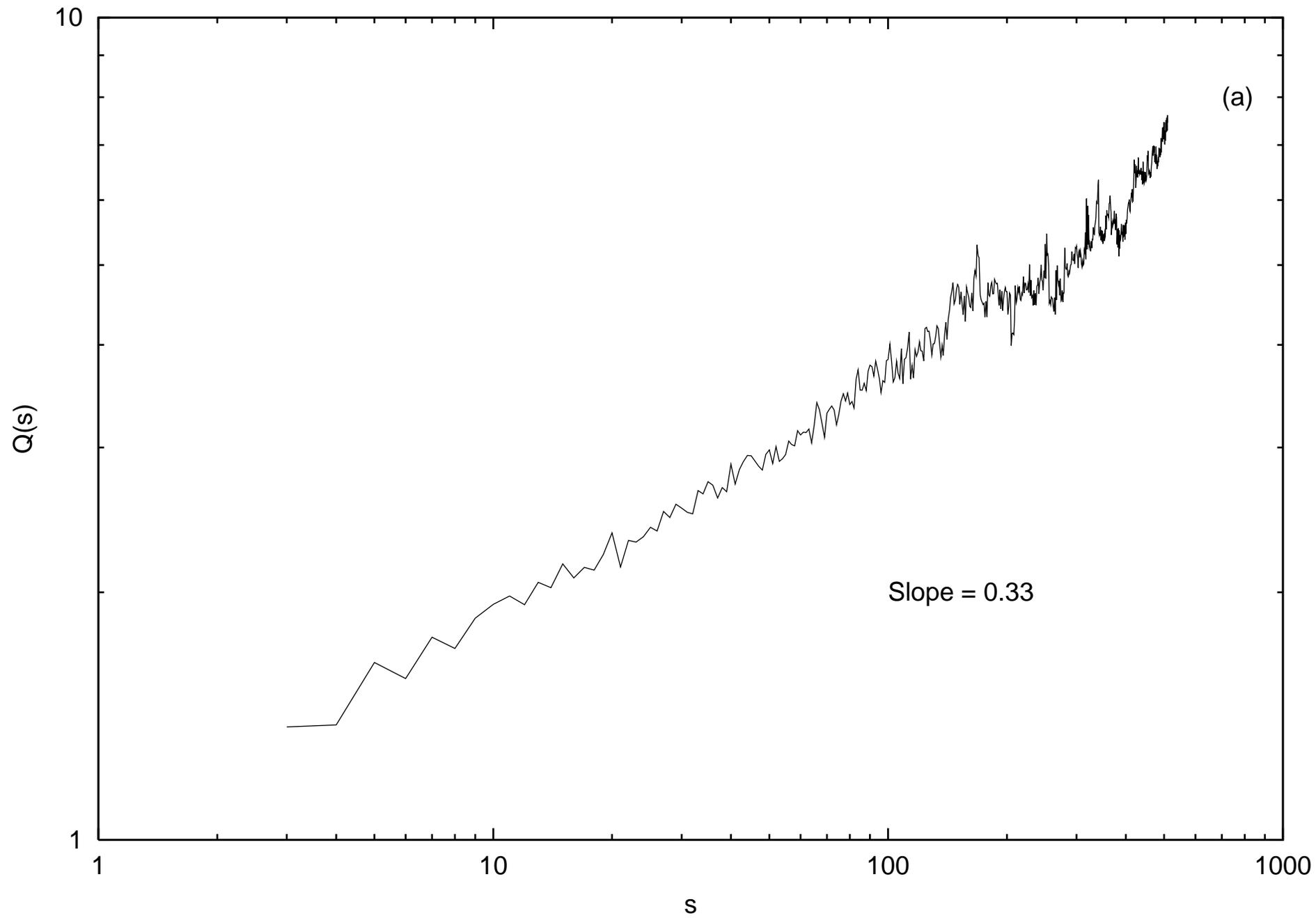
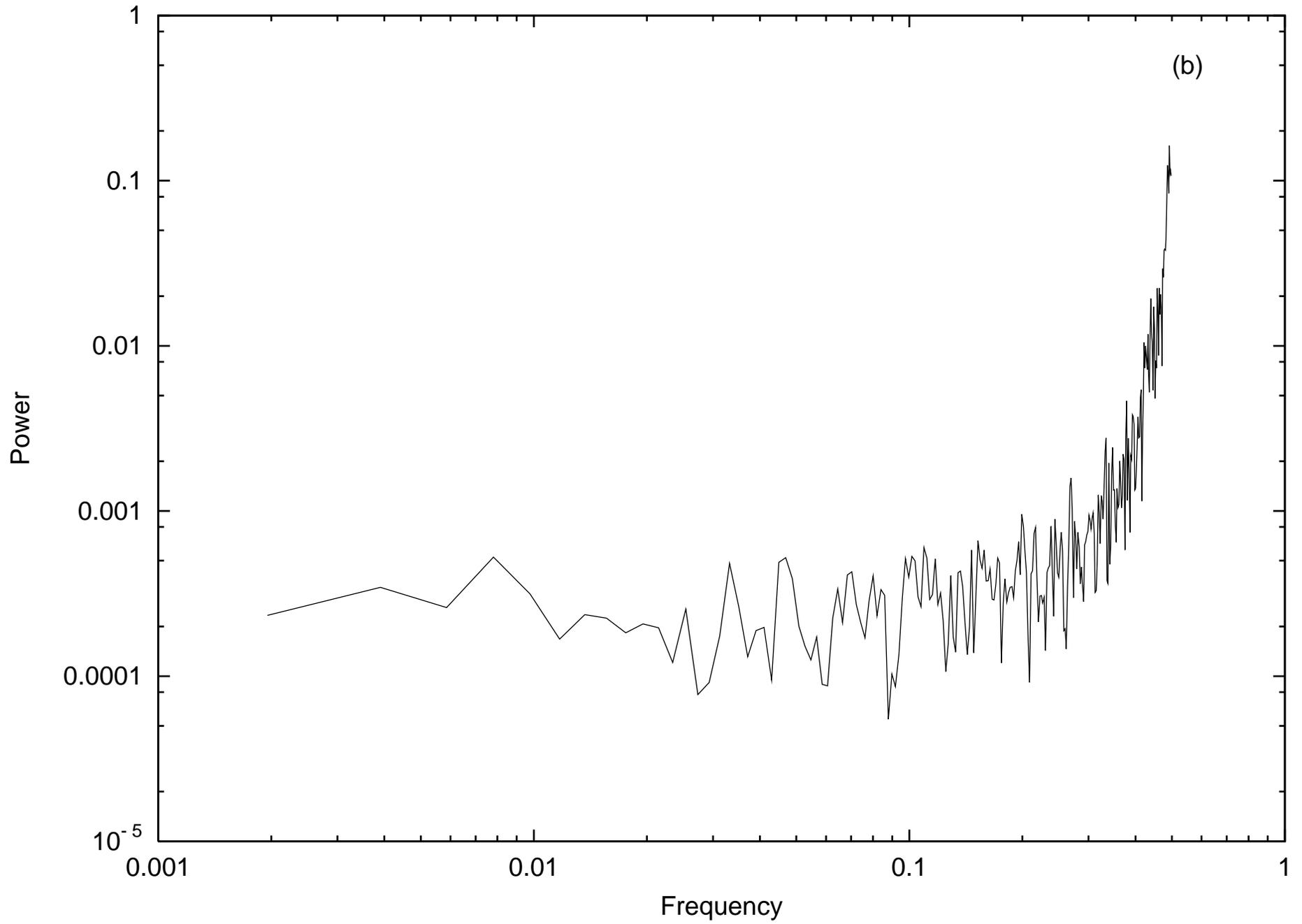
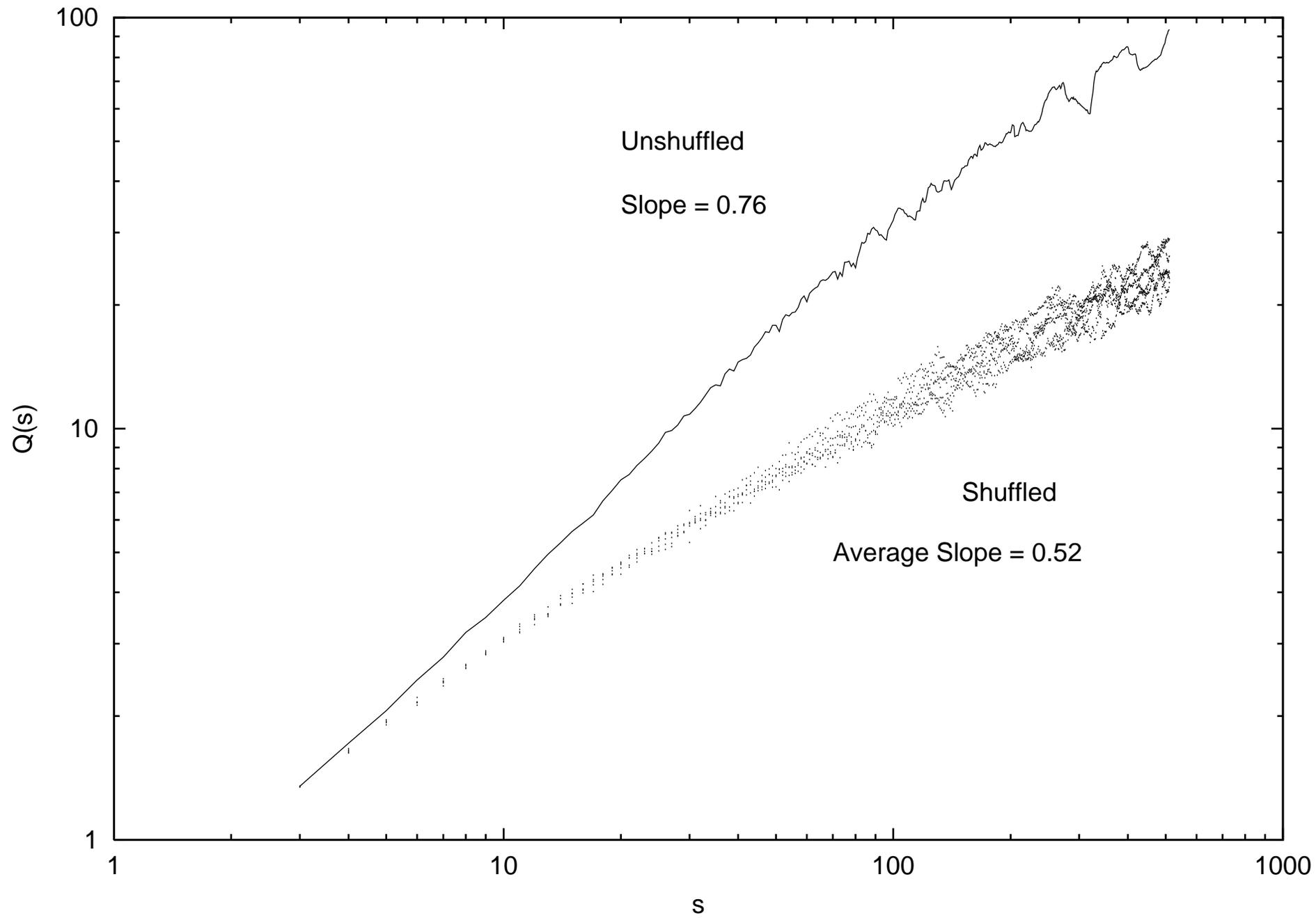
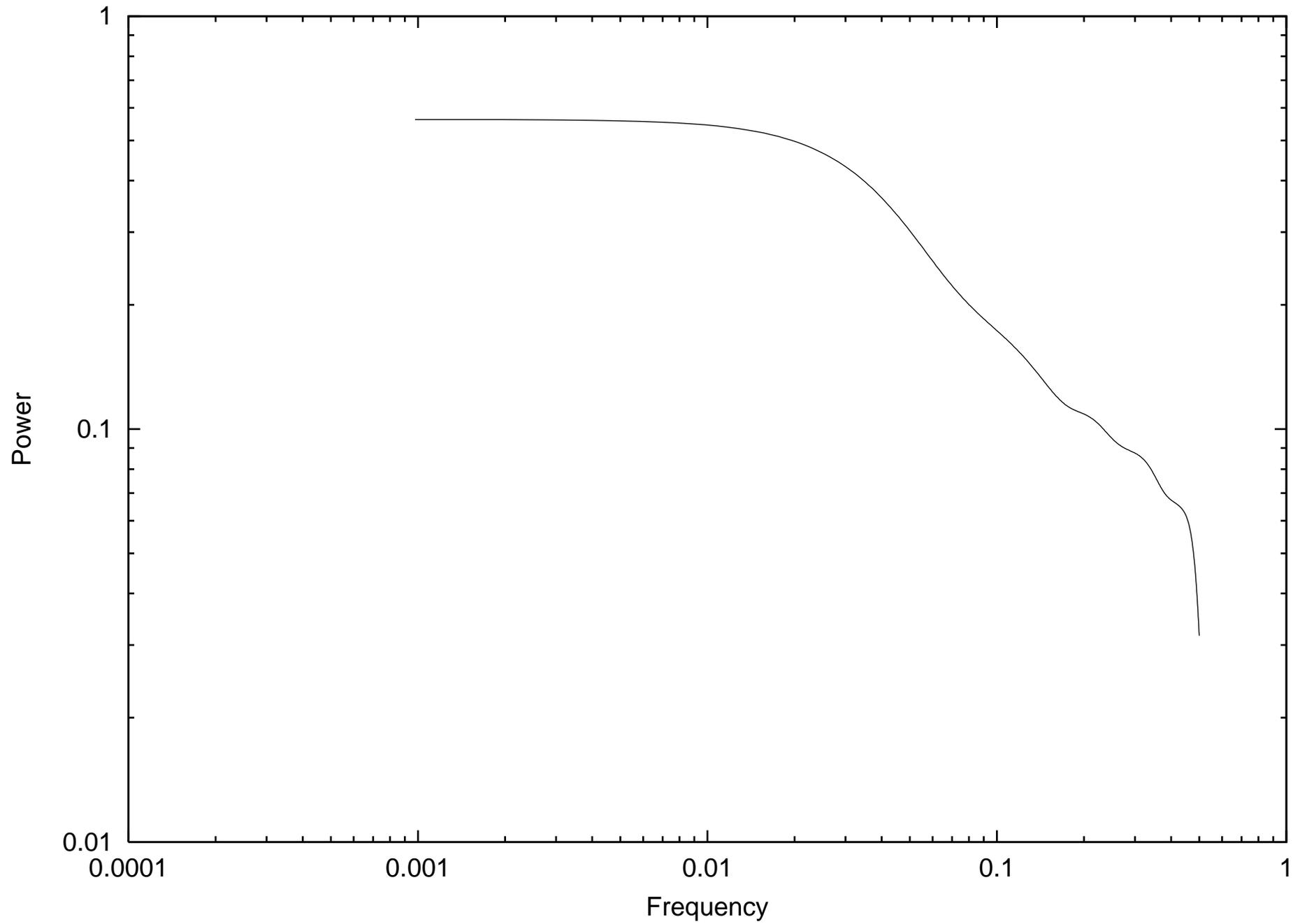
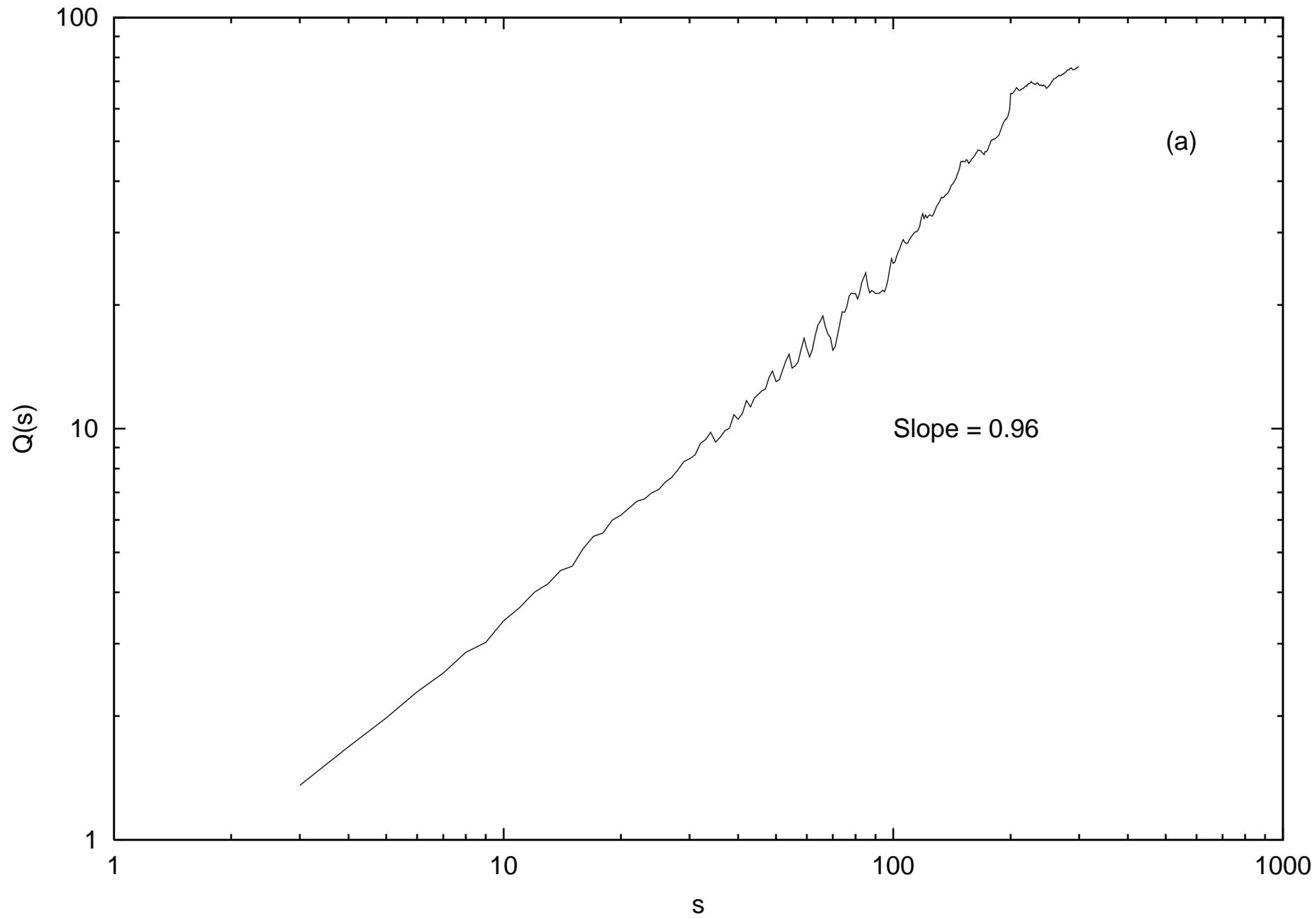
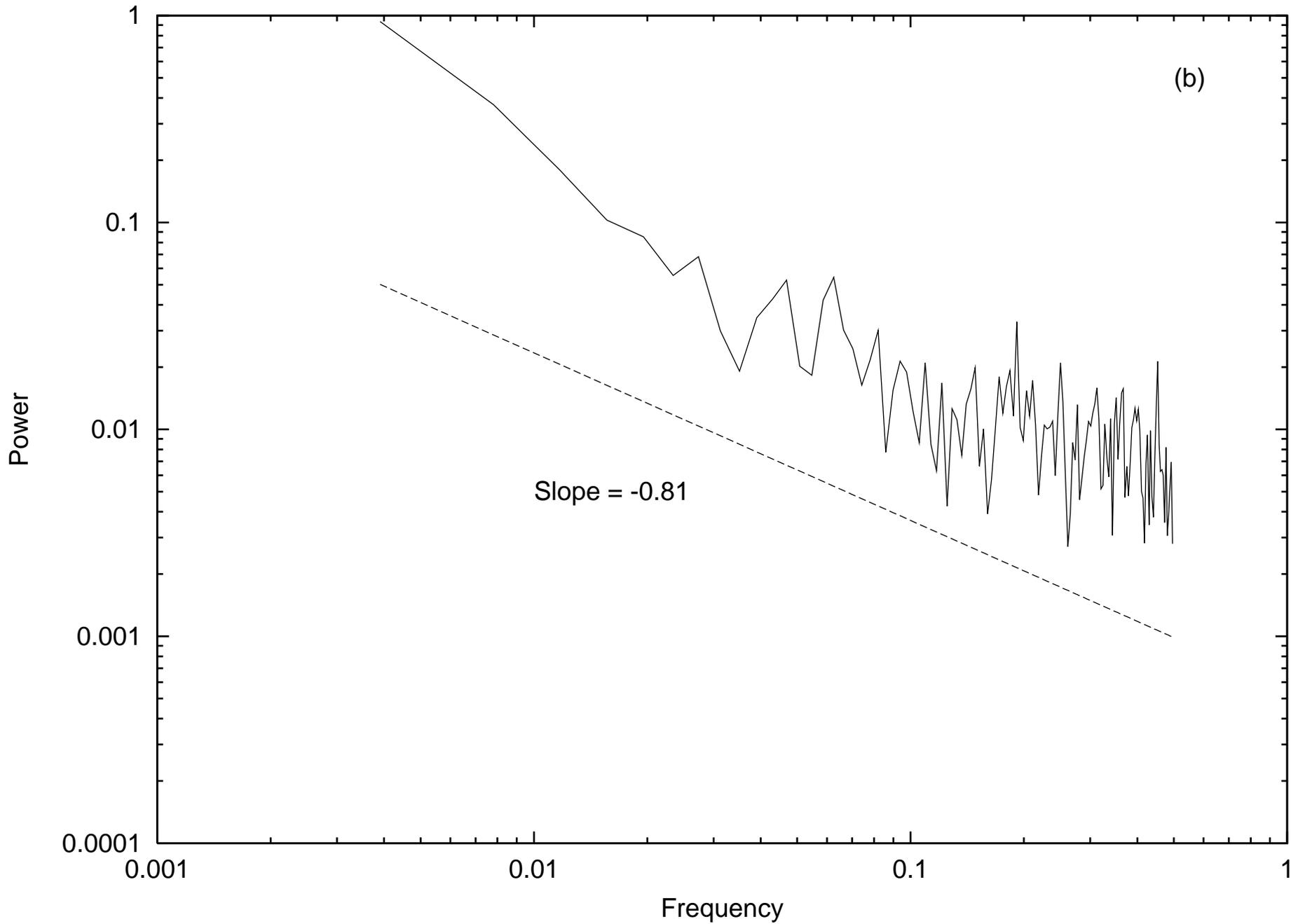
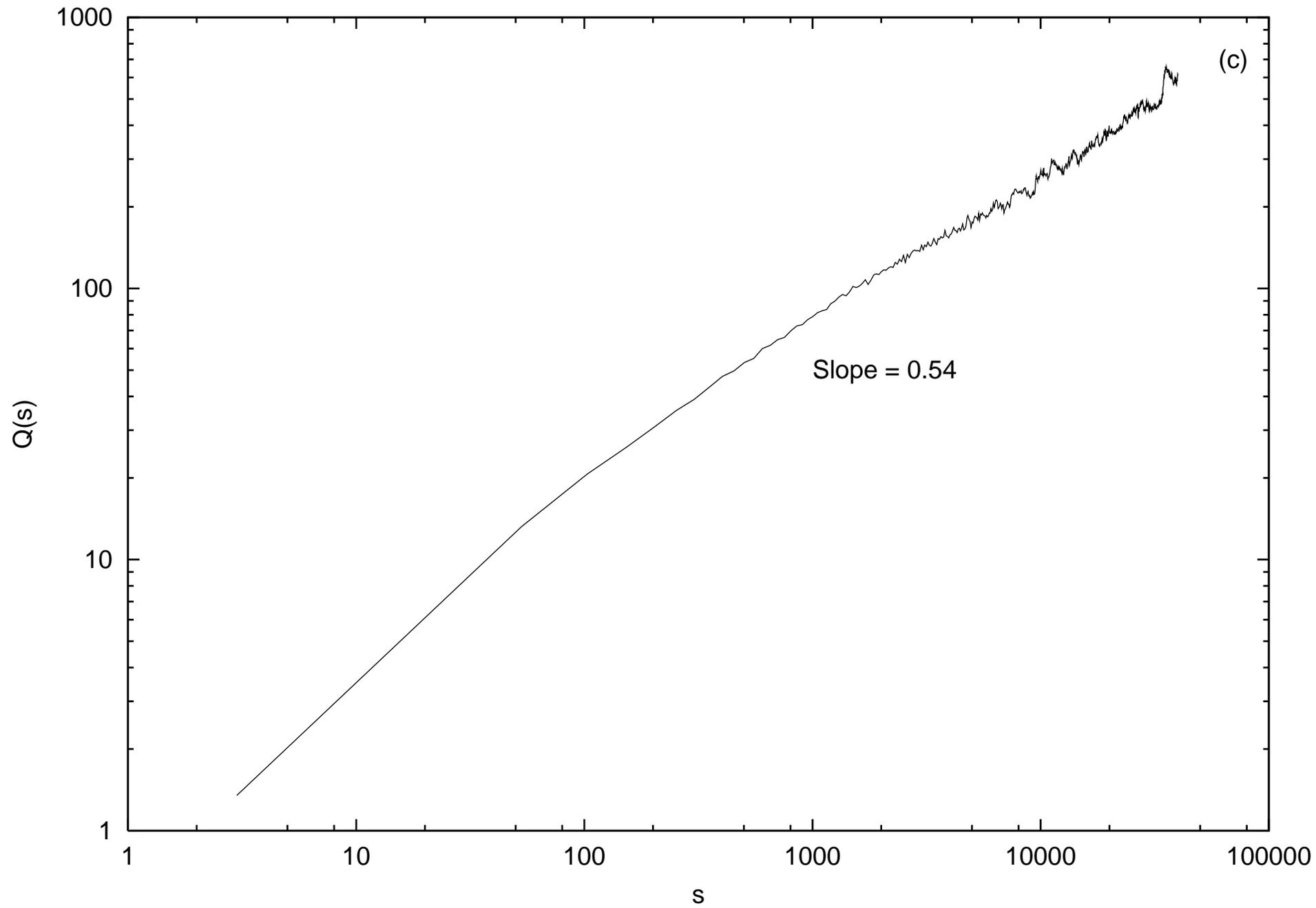
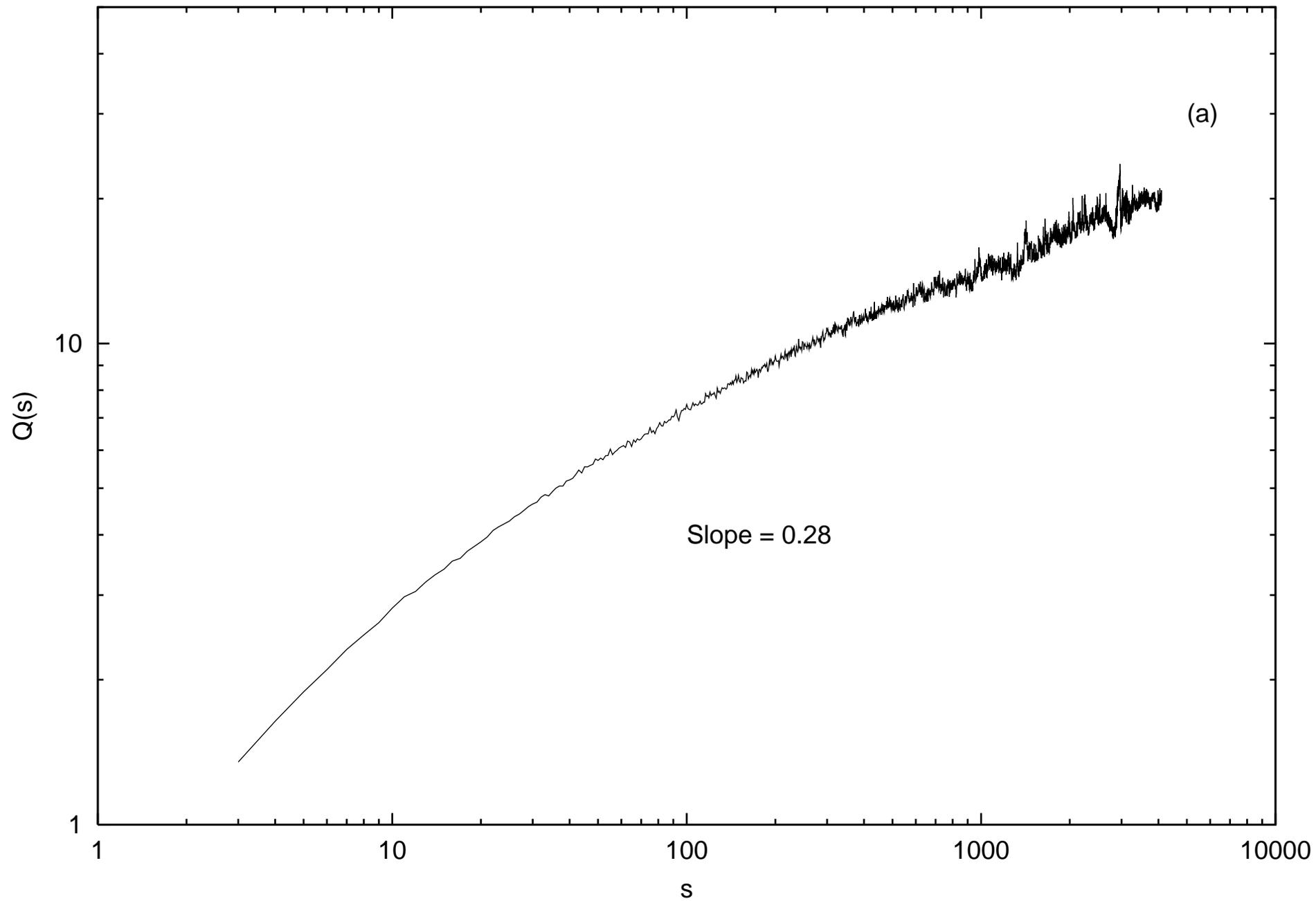
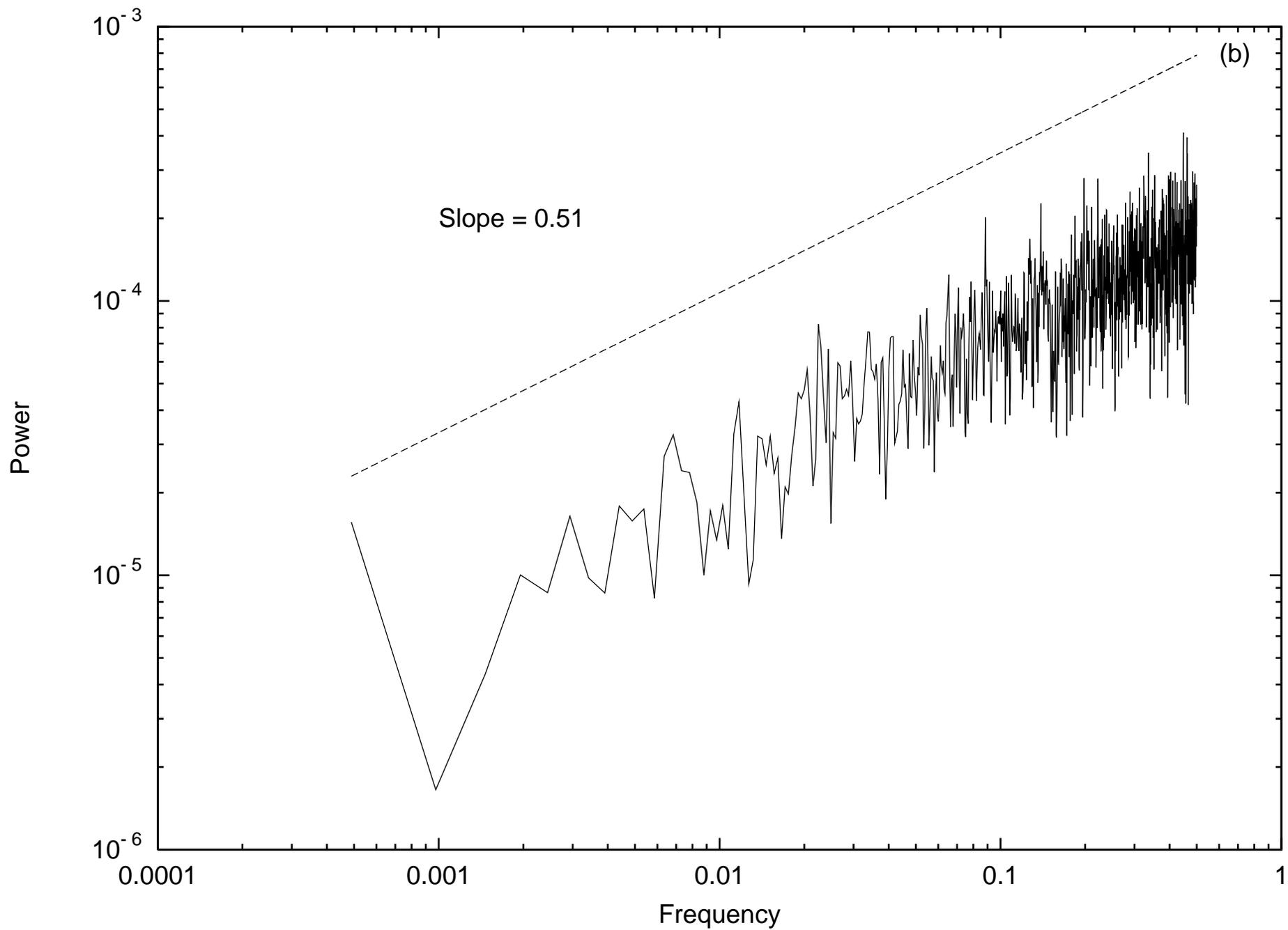
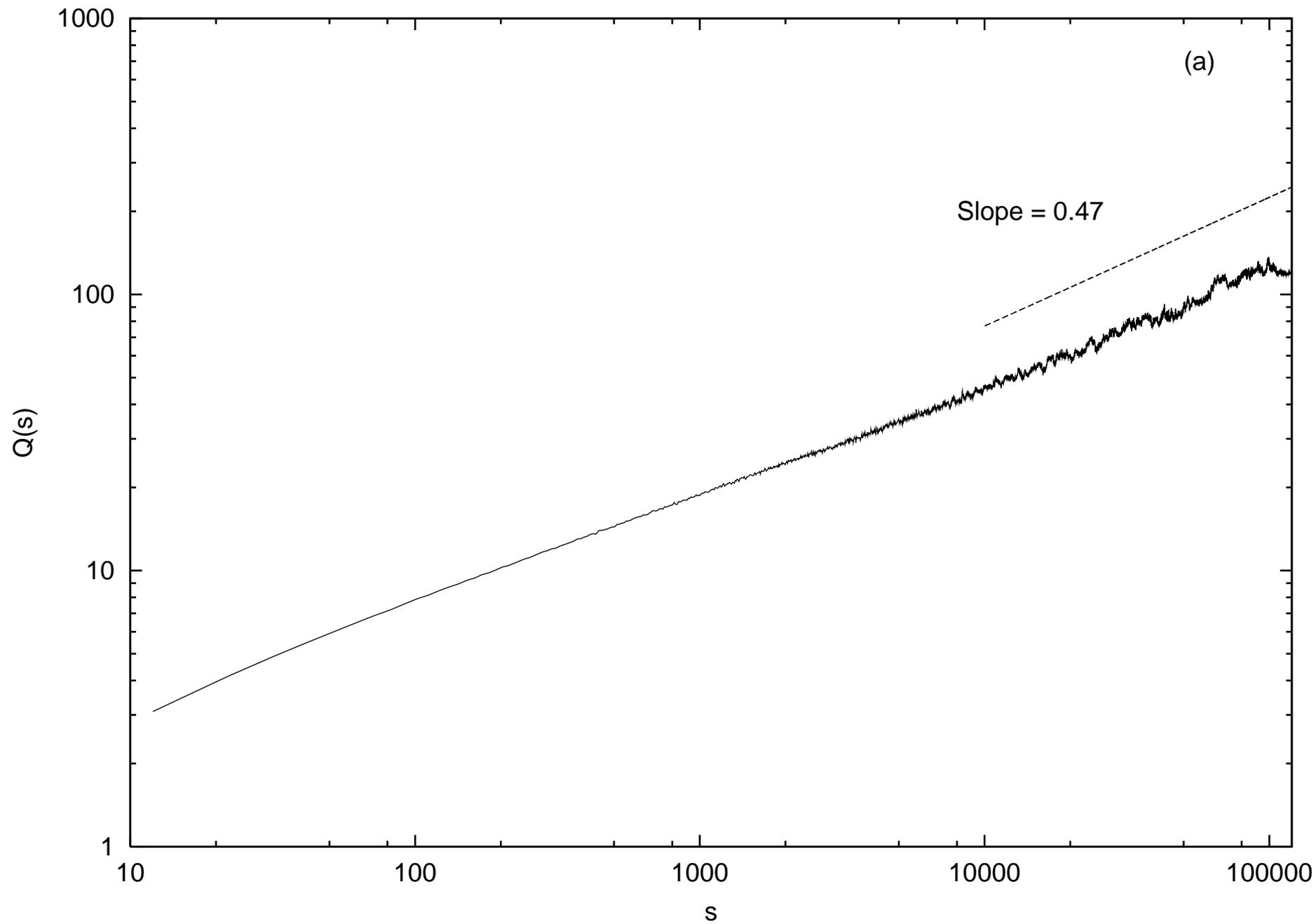
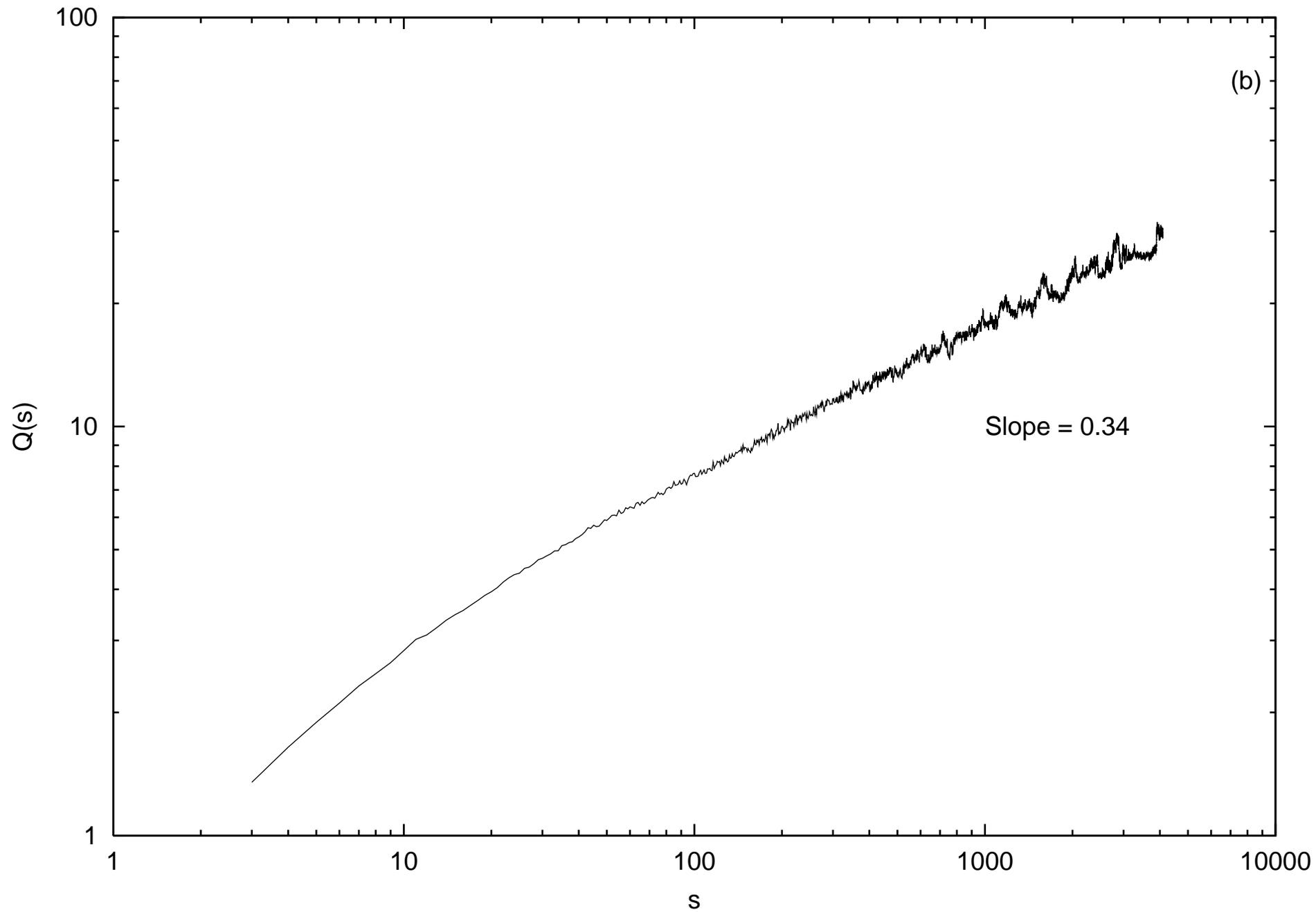
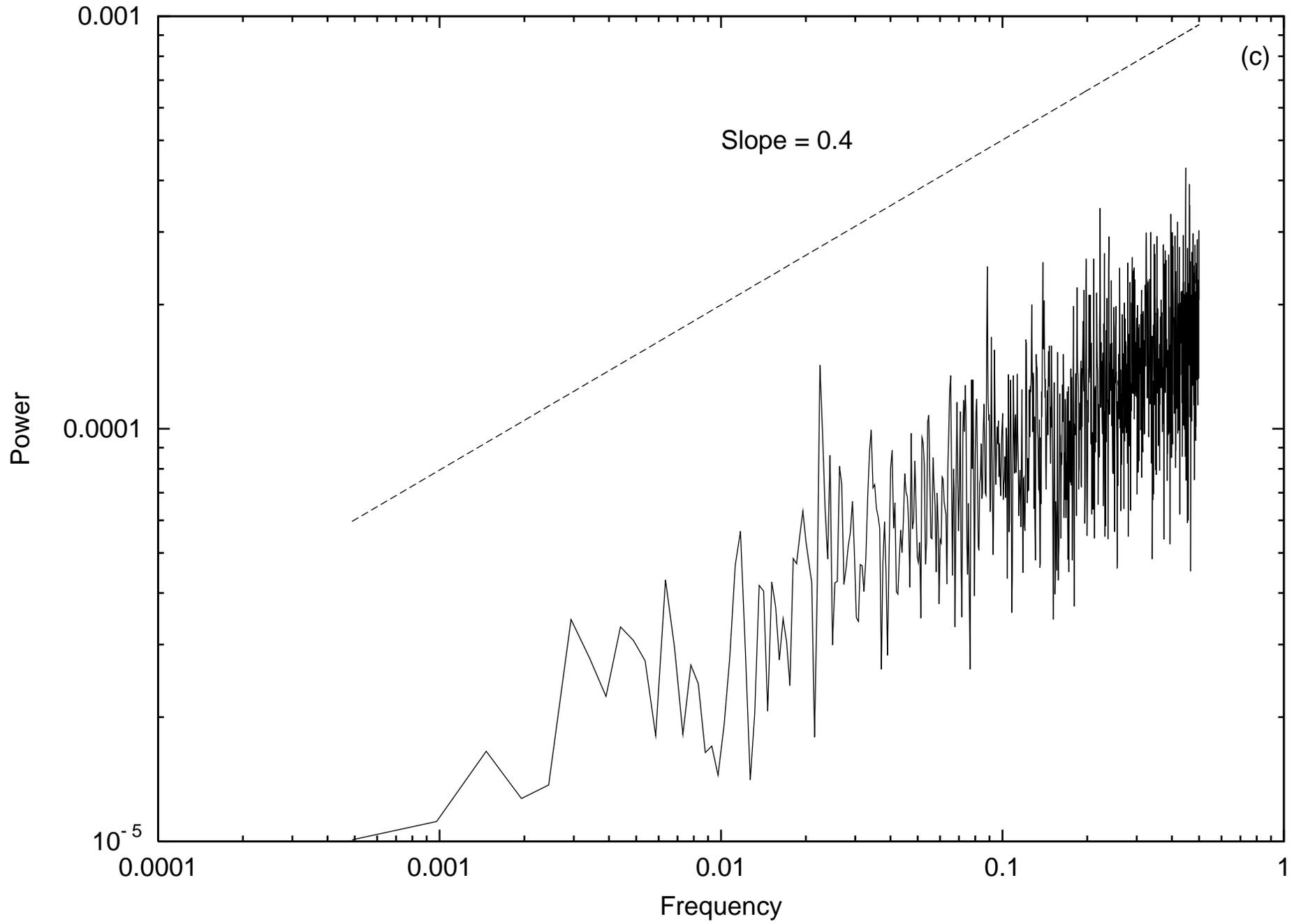
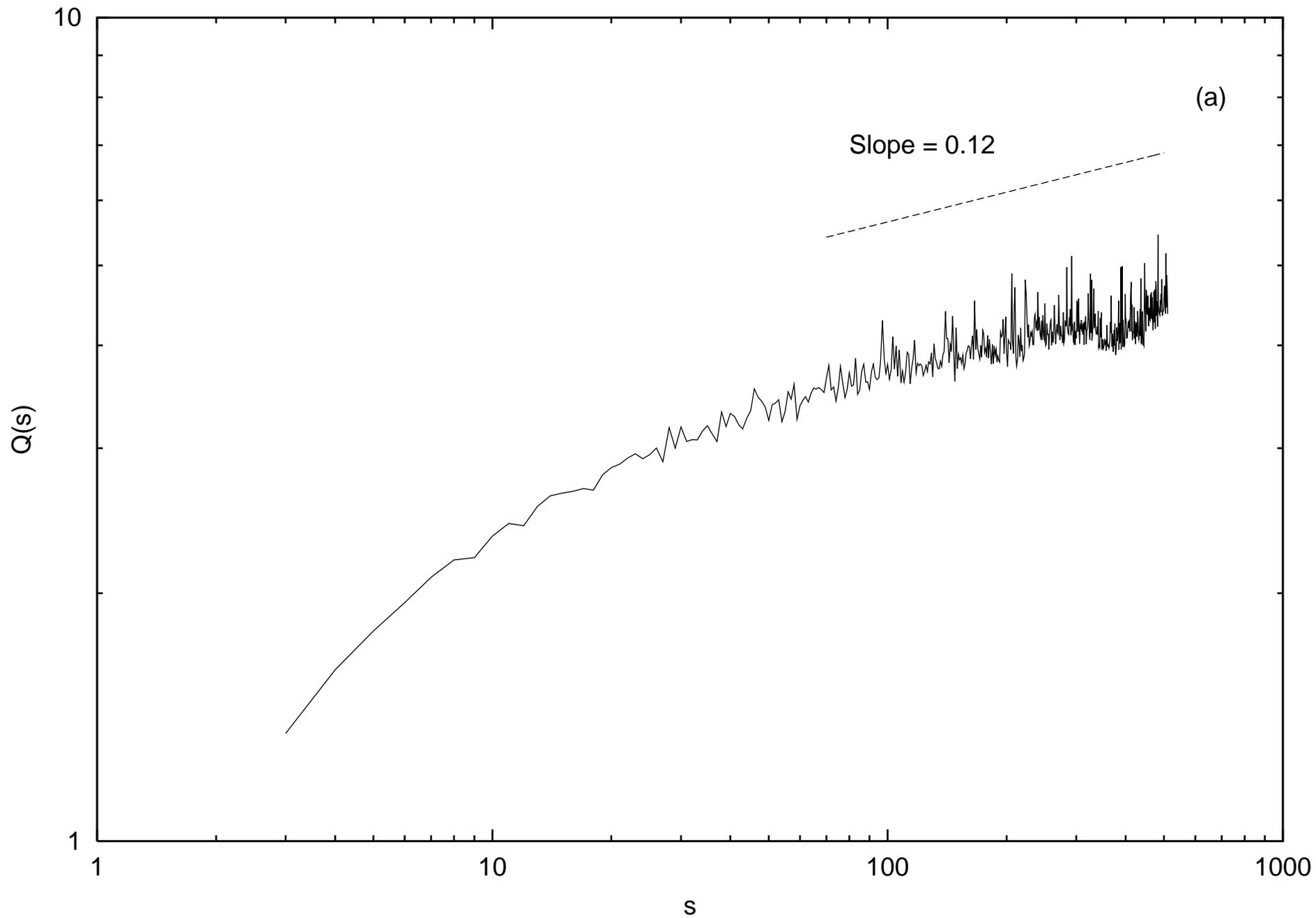
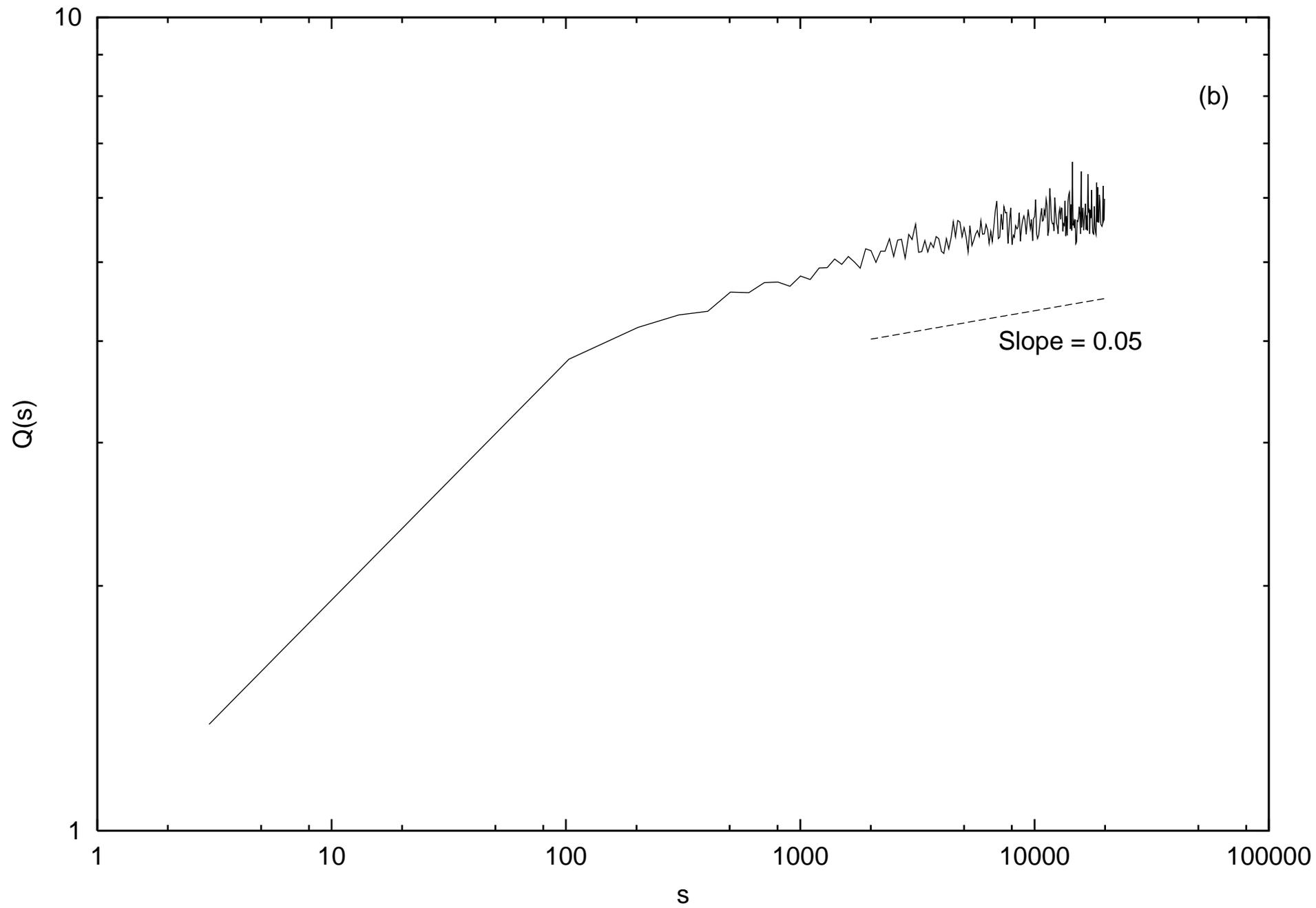
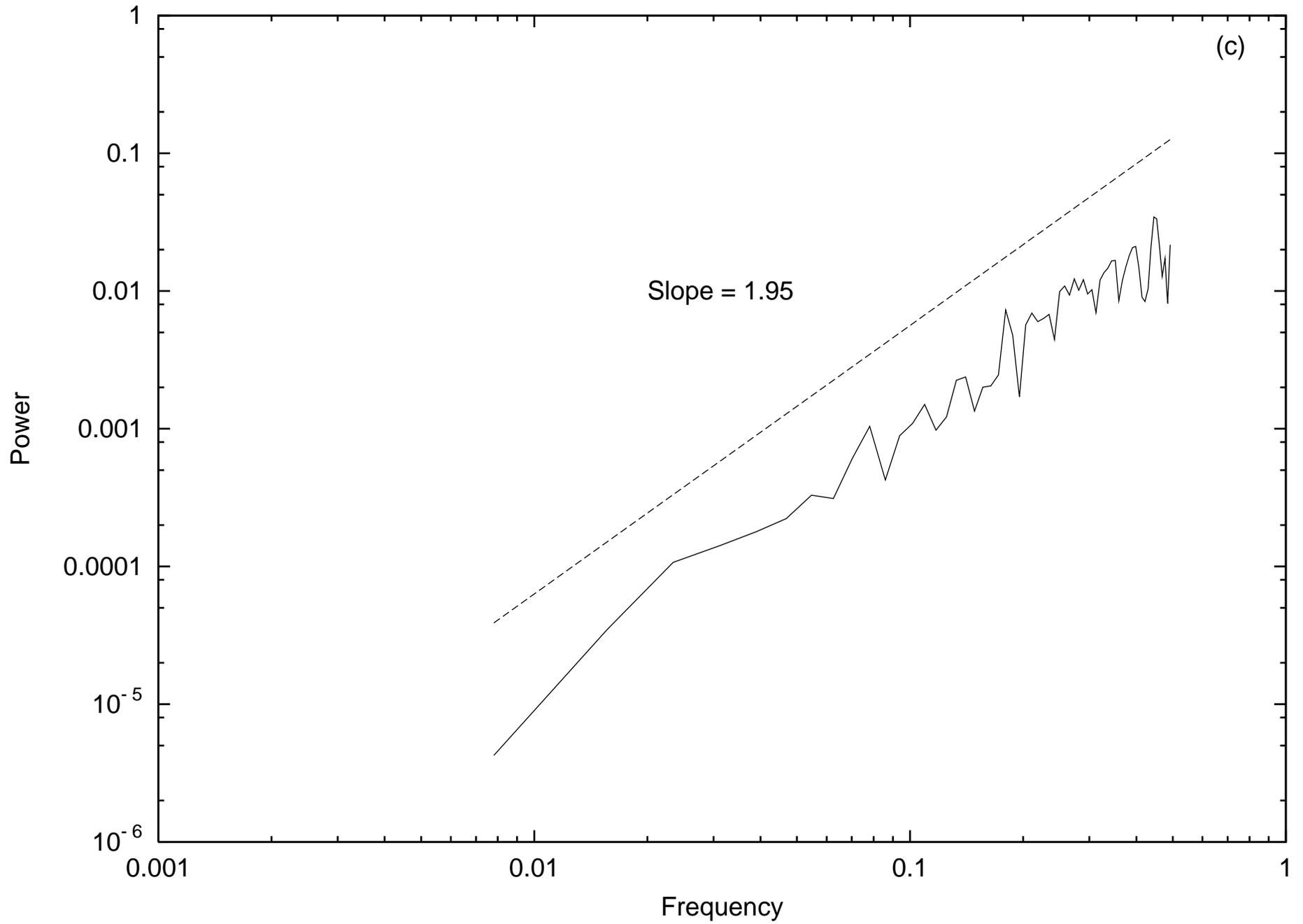
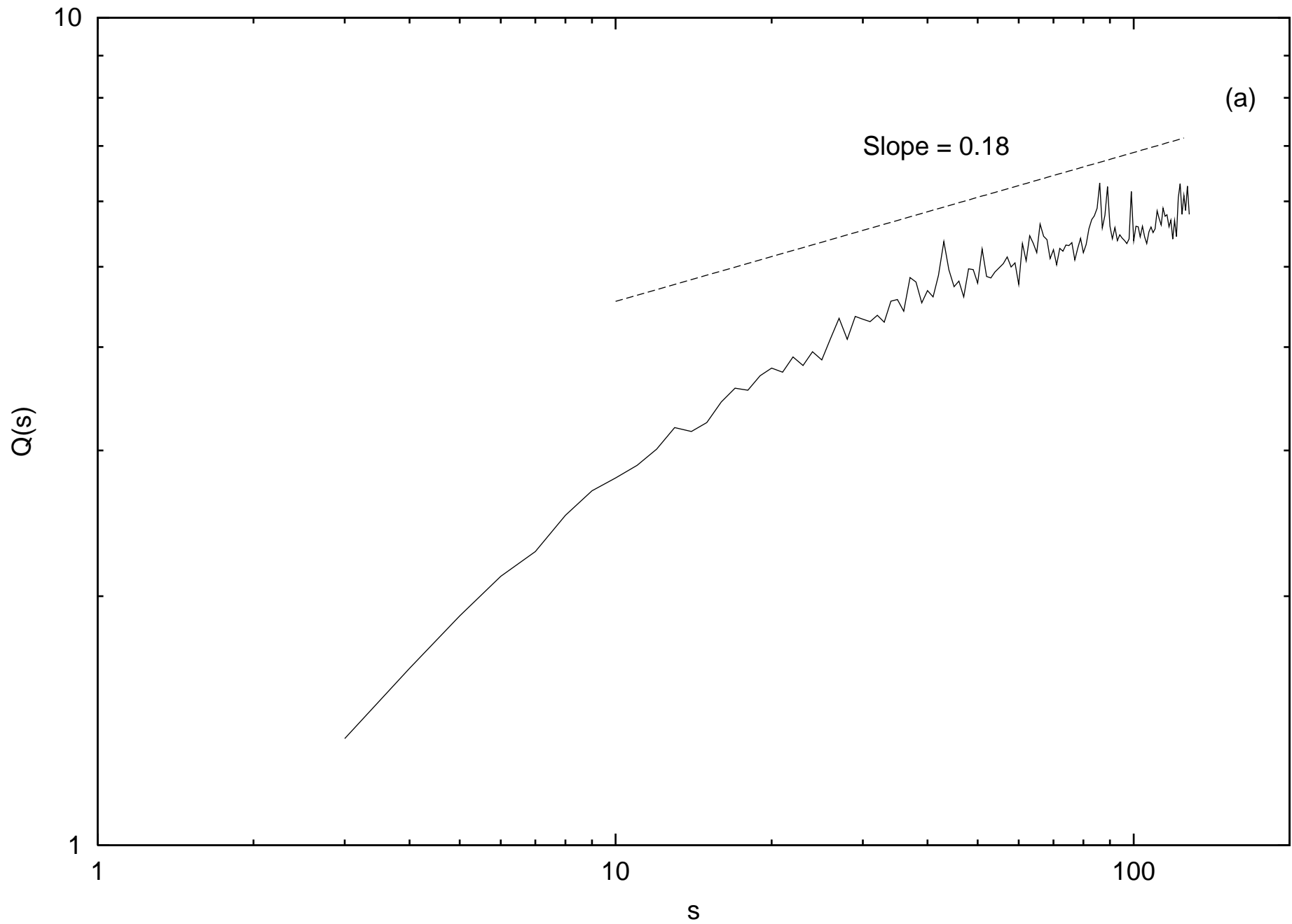
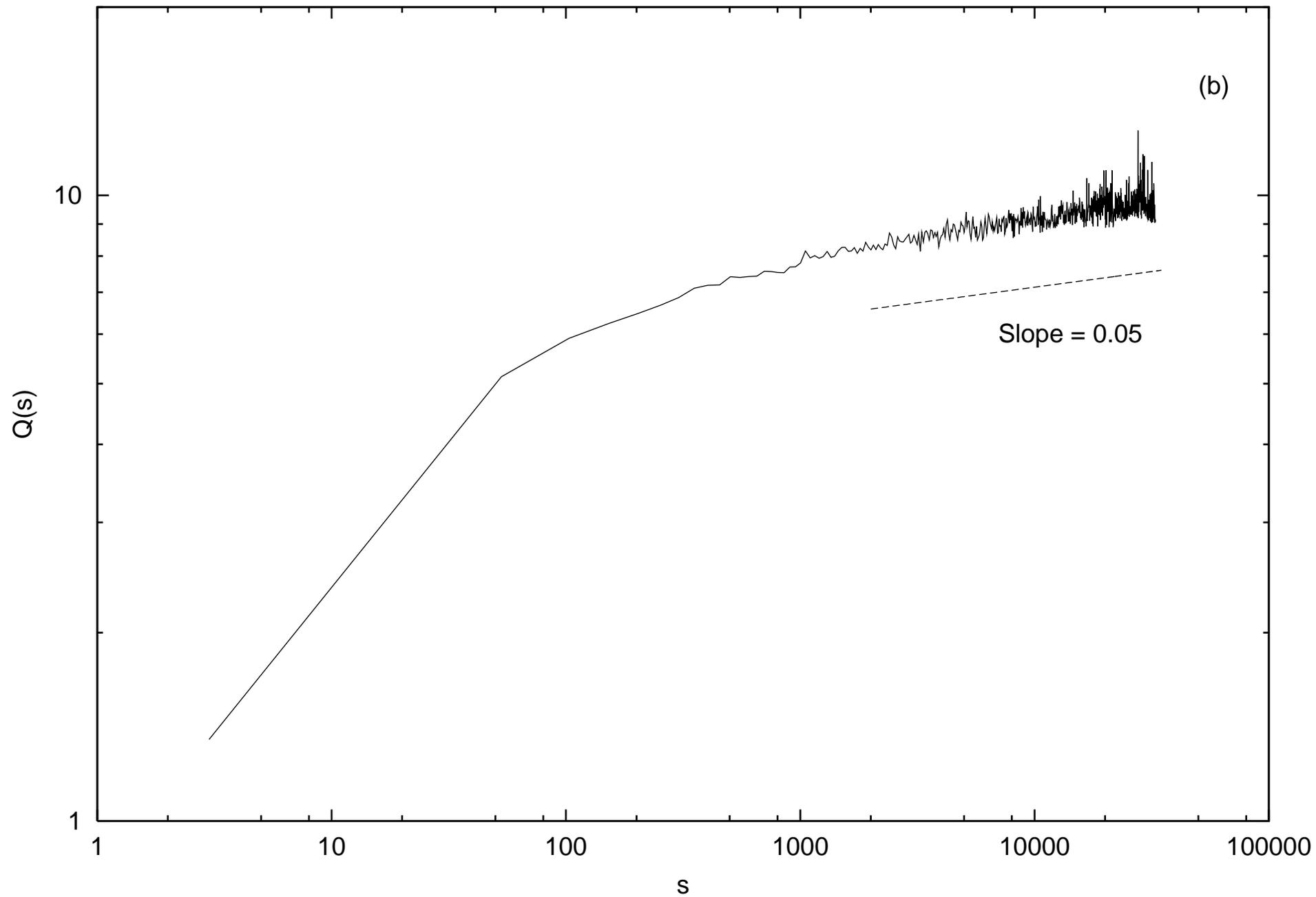
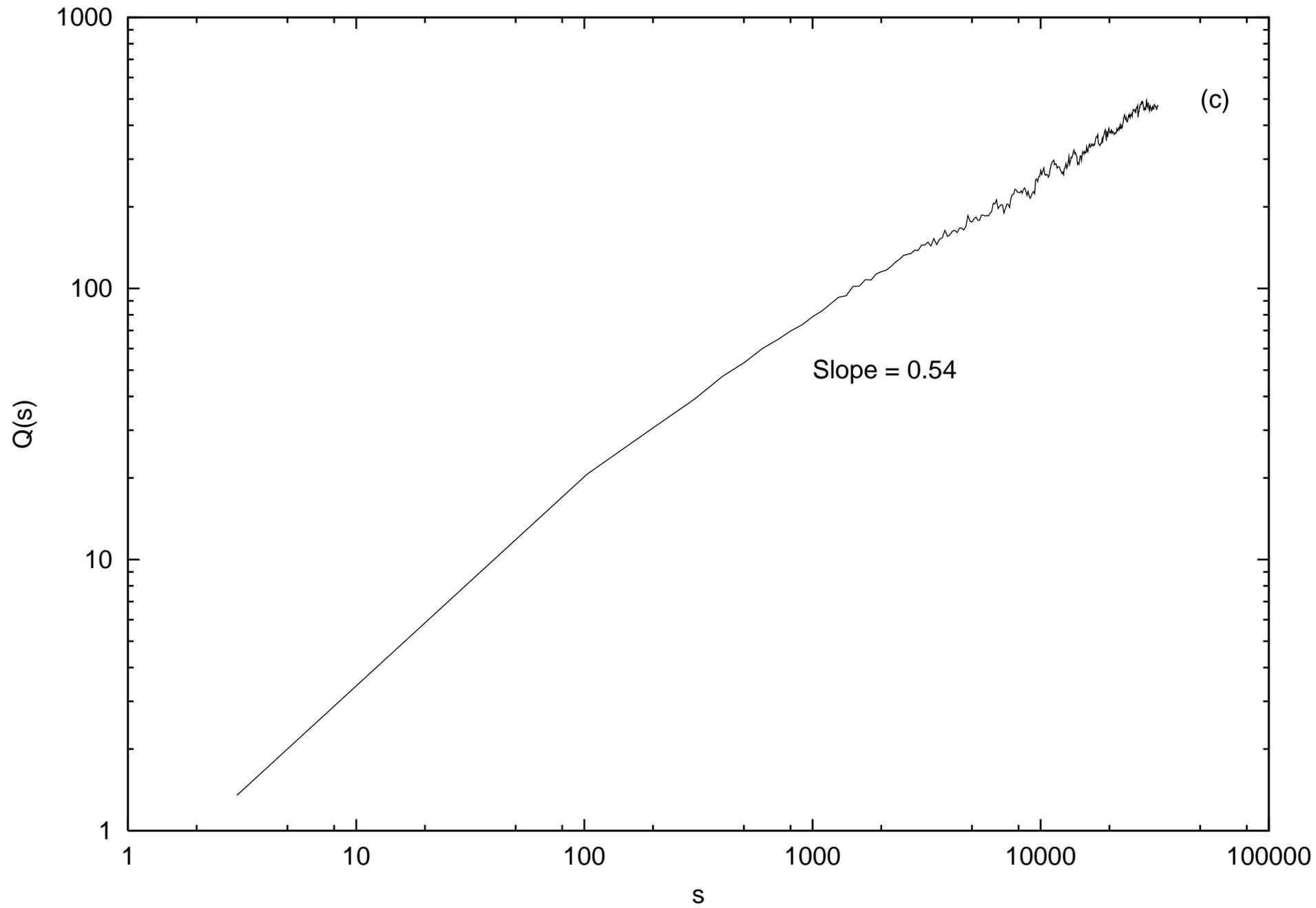
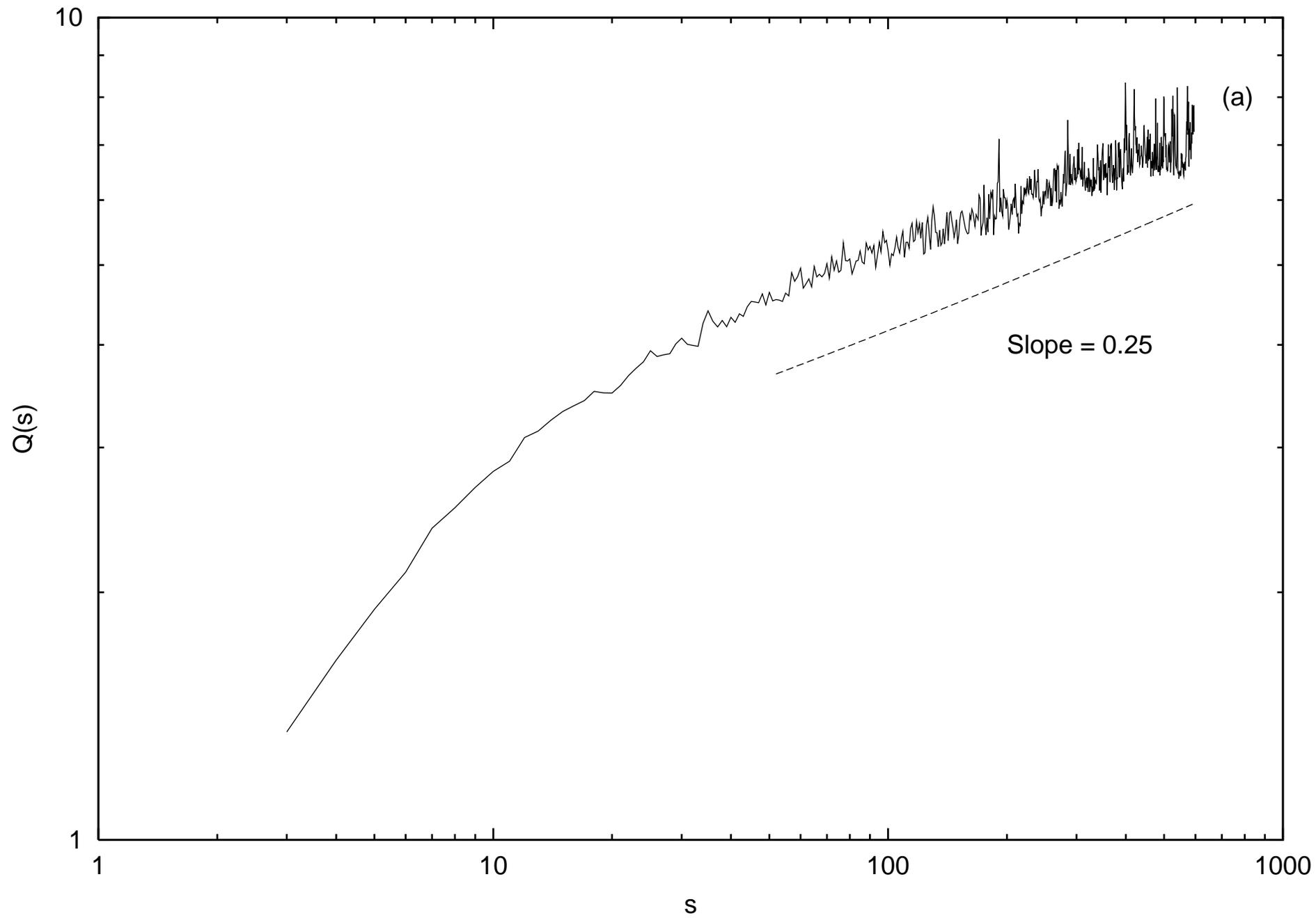
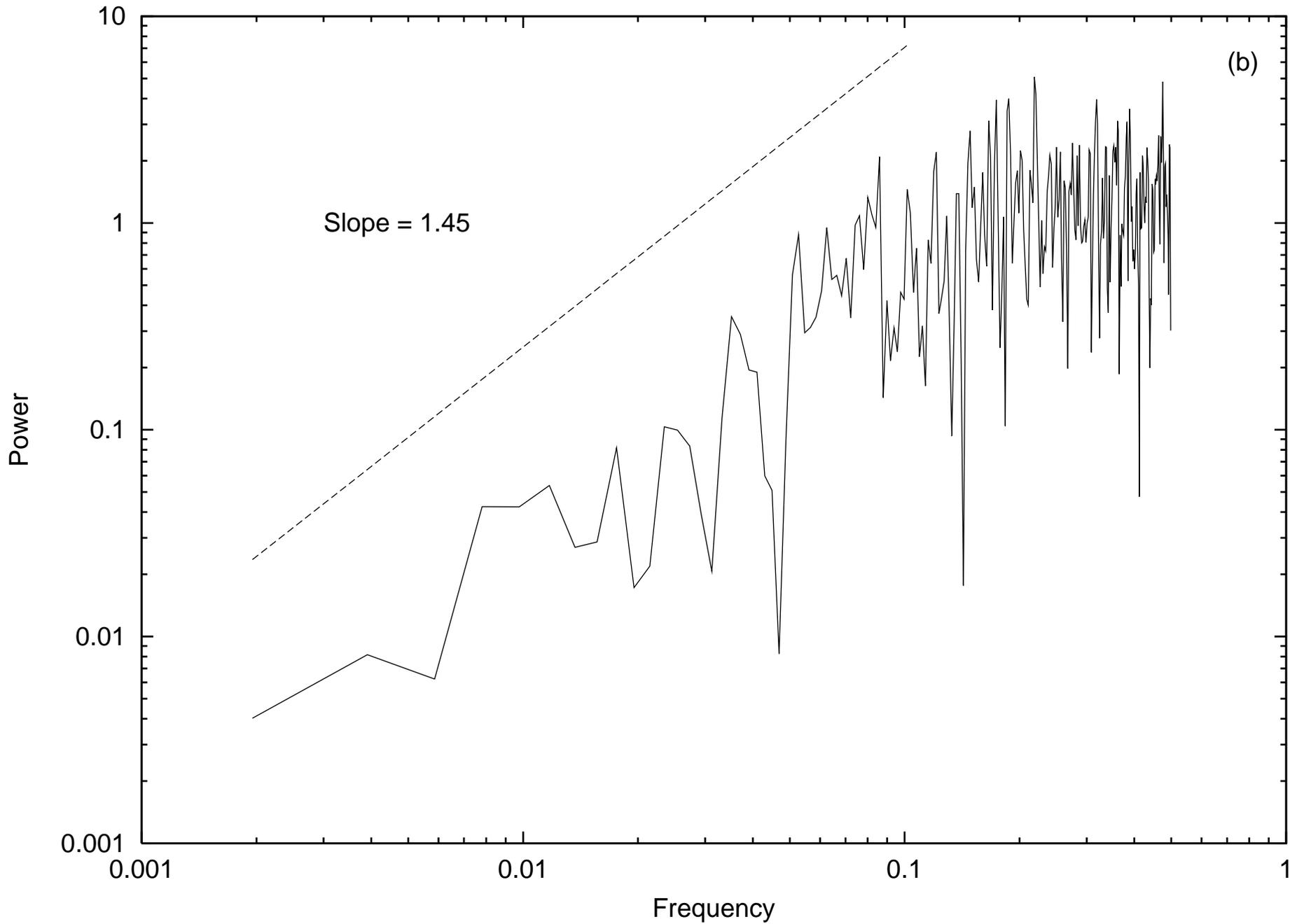
\end{document}